\documentclass[aps,twocolumn,groupedaddress,nofootinbib]{revtex4-1}

\usepackage{graphicx,color}
\usepackage{epstopdf}
\usepackage{verbatim}
\usepackage{siunitx}
\usepackage{amsmath}
\usepackage{gensymb}
\usepackage{soul}

\usepackage{xpatch}

\makeatletter
\patchcmd{\@ssect@ltx}
    {\addcontentsline{toc}{#1}{\protect\numberline{}#8}}
    {}
    {}
    {}
\makeatother

\def\thesection{\arabic{section}}

\usepackage[symbol]{footmisc}

\begin{document}

\title{Braess's paradox and programmable behaviour in microfluidic networks$^\dag$}

\author{Daniel J. Case$^{1}$, Yifan Liu$^{2}$, Istv\'an Z. Kiss$^{2}$,  Jean-R\'egis Angilella$^{3}$ \& Adilson E. Motter$^{1,4}$}
\affiliation{$^{1}$Department of Physics and Astronomy, Northwestern University, Evanston, IL 60208, USA.}
\affiliation{$^{2}$Department of Chemistry, Saint Louis University, St.\ Louis, MO 63103, USA.}
\affiliation{$^{3}$ESIX-ABTE, Universit\'e de Caen, Cherbourg 50130, France.}
\affiliation{$^{4}$Northwestern Institute on Complex Systems, Northwestern University, Evanston, IL 60208, USA.}


\begin{abstract}
Microfluidic systems are now being designed with precision to execute increasingly complex tasks. However, their operation often requires numerous external control devices due to the typically linear nature of microscale flows, which has hampered the development of integrated control mechanisms. We address this difficulty by designing microfluidic networks that exhibit a nonlinear relation between applied pressure and flow rate, which can be harnessed to switch the direction of internal flows solely by manipulating input and/or output pressures. We show that these networks exhibit an experimentally-supported fluid analog of Braess's paradox, in which closing an intermediate channel results in a higher, rather than lower, total flow rate.
The harnessed behavior is scalable and can be used to implement flow routing with multiple switches.
These findings have the potential to advance development of built-in control mechanisms in microfluidic networks, thereby facilitating the creation of portable systems that may one day be as controllable as microelectronic circuits.\\
\end{abstract}

\maketitle

{\color{white}\footnote[2]{The final version of this paper was published in Nature 574, 647 (2019), doi: 10.1038/s41586-019-1701-6.}}
\noindent
Microfluidics' promise to operate as autonomous microscale networks 
where
fluids can be transported, mixed, reacted, separated, and processed is no longer limited by experimental fabrication challenges but instead by difficulties to create built-in controls \cite{Pennathur2008,Stone2009,Perdigones2014}. The development of the modern microelectronics that form the basis of computer microprocessors was ultimately determined by the creation of integrated circuits, 
with all components  
fabricated on the same substrate. Microfluidics have already reached a level of integration in which networks with thousands of components, including control devices, are built on a single compact chip.  However, in contrast with electronic integrated circuits,  
existing on-chip fluid control devices still need to be actuated externally. For example, microfluidic circuits fabricated from flexible polydimethylsiloxane (PDMS) can now incorporate a large number of control valves, which nevertheless have to be operated using control fluids through a control layer that  
lays on top of the working fluid network \cite{Thorsen2002,Geertz2012}. As a result, microfluidics are still predominantly controlled by external hardware despite significant efforts over the past twenty years to 
develop systems with new control schemes \cite{Seker2009,Weaver2010,Tanyeri2011,Kim2012a,Li2015}.
The construction of systems that forgo the current reliance on external hardware is crucial to further the development of portable microfluidic systems for pressing applications, ranging from point-of-care diagnostics and health monitoring wearables  to analysis kits for field research \cite{Chin2012,Araci2014,Bhatia2014,Sackmann2014}. 
This requires developing
next-generation integrated circuits in which not only the control devices but also the operation of those devices is integrated on-chip. The development of such a level of integration has been fundamentally limited by the fact that, at the microscale, fluid flows tend to respond linearly to pressure changes and thus cannot be easily amplified or switched.

In this Article, we explore new physics that emerges by combining network theory and fluid mechanics to induce nonlinear behavior in microfluidics and effectively create a passive two-terminal 
 flow-switch device that is entirely operated on-chip, directly by the working fluid. 
Previous work that has achieved  built-in control capabilities (often externally actuated), including oscillatory flows \cite{Leslie2009,Mosadegh2010,Duncan2013,Duncan2015} and flow rate regulation \cite{Doh2009,Collino2013}, generally relied on flexible membranes and surfaces. Microfluidics with such flexible components require  flows with very low Reynolds numbers---a regime in which fluid inertia, and thus the only nonlinear term of the Navier-Stokes equations for 
incompressible fluids, becomes negligible. 
This has led researchers to often discount the potential effects of fluid inertia on the flows (as reviewed, for example, in Refs.~[\citenum{Stroock2002,Squires2005}]).
Recent  
work has shown, however, that inertial forces can serve as a powerful on-chip tool to manipulate microfluidic dynamics locally \cite{Amini2014,Zhang2016},
including shaping streamlines \cite{Tesar2011, Amini2013}, mixing fluids \cite{Sudarsan2006}, and directing particles \cite{DiCarlo2009,Wang2015}. 
Here, we present 
networks designed to amplify inertial effects by incorporating properties of porous media that can be used for non-local fluid routing and manipulation of output patterns.

\begin{figure} [t]
\centering
\includegraphics[width=0.99\columnwidth]{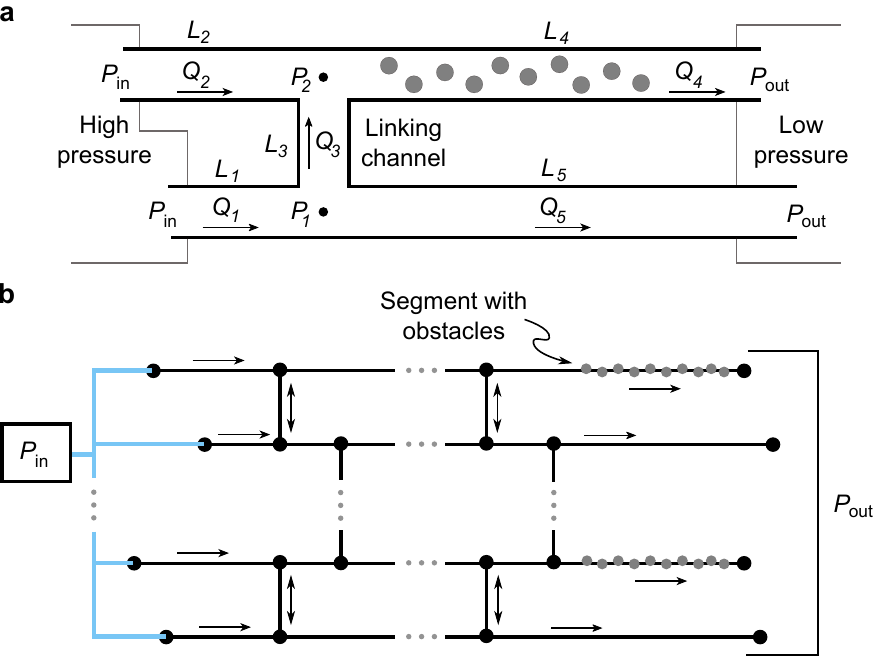}
\caption{\textbf{System schematics.} 
\textbf{a},  Microfluidic network consisting of two parallel channels, joined
by a linking channel, that connect high- and low-pressure fluid reservoirs. 
Solid gray circles represent 
stationary cylindrical obstacles. The labels denote pressures ($P$), channel lengths ($L$), and flow 
rates ($Q$), with arrows indicating the positive flow direction.
\textbf{b}, 
Generic multiswitch microfluidic network consisting of an array of parallel channels interconnected by multiple linking channels. A subset of channel segments contain cylindrical obstacles. Flow is driven through the network by a single pressure difference ($P_{\mathrm{in}}-P_{\mathrm{out}}$).
\vspace{-3mm}
}
\label{fig1} 
\end{figure}

Figure~\ref{fig1}a shows a schematic representation of a microfluidic system with the fundamental network structure we consider.
It consists of five segments
arranged as two parallel channels connected by a linking channel, where the inlets are kept at a common pressure $P_{\mathrm{in}}$
and the outlets are held  at a common lower pressure $P_{\mathrm{out}}$.  One of the outlet channels is modified to generate a 
nonlinear pressure-flow relationship, which is achieved  by introducing an array of cylindrical obstacles.
Our principal results are supported by theory, simulations, and experiments, and they show that we can:
1) induce a flow direction {\it switch} through the linking channel solely by varying the pressure difference between the inlets and outlets;
2) identify a pressure difference above which the total flow rate
between the inlets and outlets {\it increases} upon closing the linking channel.
We also predict negative conductance transitions when the linking channel is equipped with an offset fluidic diode, which are analogous to
non-monotonic pressure-flow relations
previously observed using flexible diaphragm valves \cite{Xia2014}.
The counter-intuitive behavior described in (2)
is formally equivalent to the so-called Braess paradox originally established for traffic networks \cite{Braess1968,Braess2005}, where closing a shortcut road has the possible effect of increasing net traffic flow. 
We demonstrate integration of the flow switch described in (1)
by considering larger microfluidic networks, as illustrated in Fig.~\ref{fig1}b, which incorporate multiple linking channels and are thus capable of exhibiting multiple flow switches. Flows through these networks are driven by a single pressure difference and yet can be designed to exhibit a variety of flow states by programming the pressure at which each flow switch occurs.

\subsection*{System design and nonlinearity}
We consider conditions under which all channel segments have the same width $w$, the working fluid is water, and all surfaces (including obstacles) have no-slip boundaries. We assume, without loss of generality, that the pressure $P_{\mathrm{out}}$ at the outlets is zero, and consider 
scenarios in which either the static or total pressure is controlled at the inlets (Methods).
We examine two network configurations of the system in Fig.~\ref{fig1}a: the {\it connected configuration}, in which the two parallel channels are allowed to exchange fluid through the linking channel;  and the {\it disconnected  configuration}, in which the linking channel is closed or removed. In our theoretical analysis and simulations, the flows are assumed to be two dimensional, yet the main results carry over to three dimensions, as verified in our experiments.

\begin{figure}[tb] 
\includegraphics[width=\columnwidth]{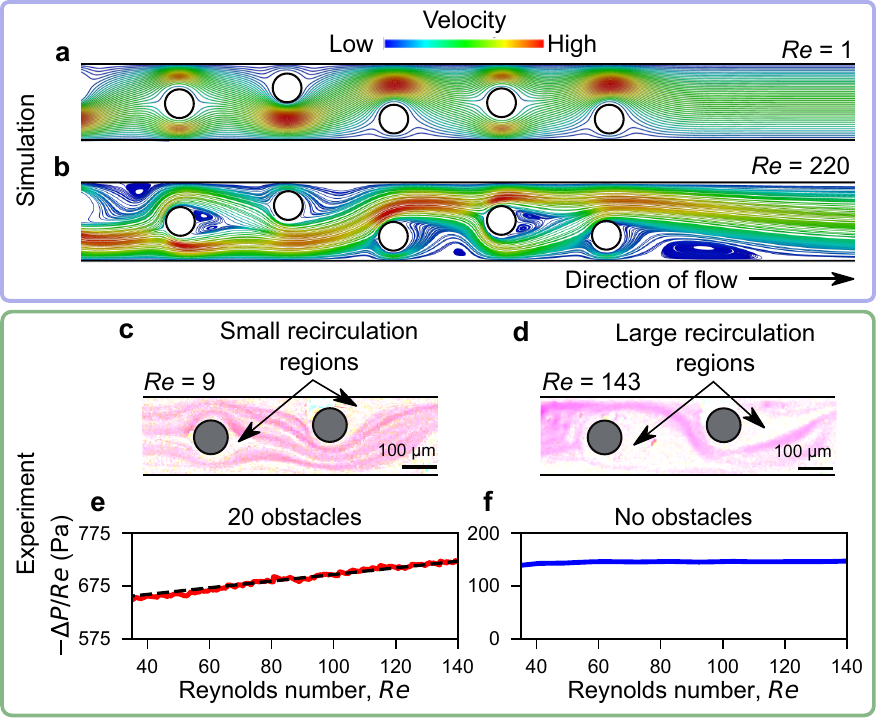}
\caption{\textbf{Development of nonlinear flow.} \textbf{a}, \textbf{b}, Simulated flow in a channel with obstacles (white circles), showing no recirculation for low $Re$ (\textbf{a}) and significant recirculation near the obstacles for larger $Re$ (\textbf{b}). \textbf{c}, \textbf{d}, Experimentally observed flows around the obstacles (grey circles), visualized using
pictures of fluorescent particles (shown in pink). The particle tracks trace the underlying flow structure, confirming the development of recirculation regions (white areas) as $Re$ is increased from low (\textbf{c}) to moderate values (\textbf{d}).
\textbf{e}, \textbf{f}, Experimentally measured relation between pressure loss and $Re$  for a channel with (\textbf{e}, red curve) and without (\textbf{f}, blue curve) obstacles. The dashed line in \textbf{e} is a reference to guide the eye and indicates an approximately quadratic relation between pressure loss and flow rate. 
}
\label{fig2}
\end{figure}

For a straight microfluidic channel of length $L\gg w$ without obstacles,
an approximate steady-state solution of the Navier-Stokes equations in two dimensions yields a linear relation between the total volumetric flow rate per unit depth $Q$
and the pressure drop $\Delta P$ along the channel,
\begin{equation} \label{eq1}
-\Delta P = \frac{12\mu L}{w^3} Q,
\end{equation}
where $\mu$ is the dynamic viscosity of the fluid.
To induce deviations from this linear regime, we consider the effect of introducing multiple stationary obstacles in the channel.
Figure~\ref{fig2}a,b shows simulations of the Navier-Stokes equations for a channel with ten cylindrical obstacles of radius $r = w/5$ 
(Methods).
We observe recirculation regions forming near the obstacles for sufficiently large Reynolds number $Re\equiv 2\rho Q/\mu$, where $\rho$ is the fluid density. 
The recirculation regions first appear for $Re$ of order $10$, and their number and size depend on $Re$.
These localized structures
are hallmarks of fluid inertia effects (and thereby of nonlinearity).
We investigate how fluid inertia effects compound to impact the total flow rate by performing simulations across moderate values of $Re$ 
when different numbers of obstacles are present.
We find that a nonlinear relation between the pressure drop $\Delta P$ and flow rate $Q=\mu Re/2\rho$ emerges as soon as obstacles are introduced and that the nonlinearity becomes more pronounced as
the number of obstacles is increased (Supplementary Information, section~\ref{S3.1} and  Fig.~\ref{fig5SI}).

\begin{figure}[t] 
\centering
\includegraphics[width=0.99\columnwidth]{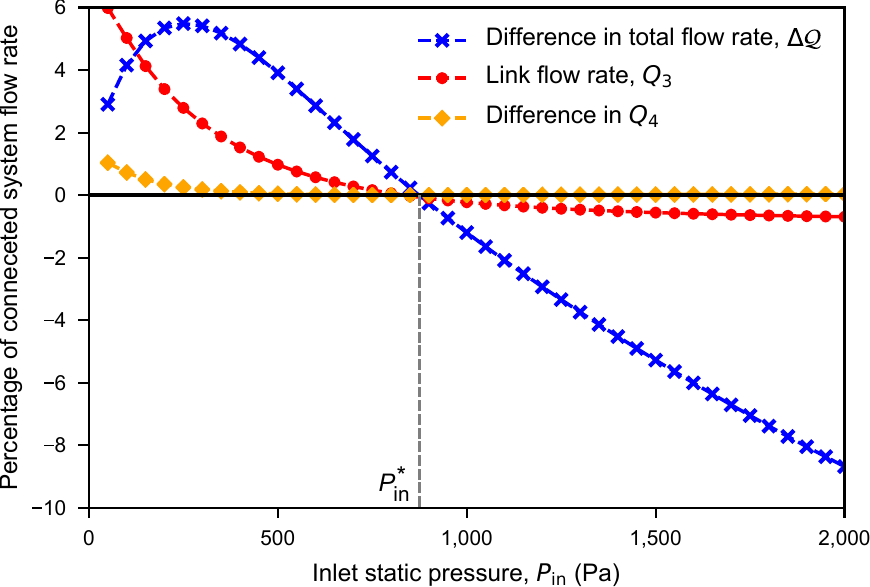}
\caption{\textbf{Braess's paradox and flow switching.} Simulation results for the connected and disconnected configurations of the system for a range of inlet pressures $P_{\mathrm{in}}$.
The flow rates are presented as a percentage of the total flow rate through the connected system, $Q_C$, where we adopt the sign  convention for the flow directions as defined in Fig.~\ref{fig1}a.
The flow through the linking channel switches direction at the critical pressure $P_{\mathrm{in}}=P_{\mathrm{in}}^*$,  which coincides with the onset of negative $\Delta \mathcal{Q}$ that marks the occurrence of Braess's paradox.
}
\label{fig3}
\end{figure}

The nonlinearity we observe in the relation between $\Delta P$ and $Q$
conforms to the well-known Forchheimer effect in porous media, which characterizes flow through many interconnected microchannels where local inertial effects at the points of interconnection are non-negligible, even for creeping flow \cite{Rojas1998,Andrade1999,Fourar2004}. 
We  use the Forchheimer equation to derive a relation between $\Delta P$ and $Re$ for the channel with obstacles, given by
\begin{equation} \label{eq2}
-\Delta P = \frac{\alpha \mu^2 L}{2\rho w}Re + \frac{\beta \mu^2 L}{4\rho w^2}Re^2,
\end{equation}
where $\alpha$ is the reciprocal permeability and $\beta$ is the non-Darcy flow coefficient, both depending solely on the system geometry (Methods).

The physical mechanism giving rise to this nonlinearity is the increase in flow recirculation and velocity gradients for larger $Re$, as evidenced in Fig.~\ref{fig2}a,b for $Re=1$ and $220$.
To test the impact of the inertial effects in realistic systems, we perform experiments using microchannels fabricated
from 
stiff 
PDMS (hardened by curing). 
Figure~\ref{fig2}c,d shows experimental evidence
of the increase in the number and size of 
the recirculation regions with $Re$, in agreement with our simulations.
An approximately linear relation between $-\Delta P/Re$ and $Re$ and thus an approximately quadratic relation between $-\Delta P$ and $Q$ for a channel containing twenty obstacles is shown in Fig.~\ref{fig2}e, which contrasts with the constant relation measured for a channel without obstacles in Fig.~\ref{fig2}f.

\begin{figure*}[htb!] 
\centering
  \includegraphics[width=0.99\textwidth]{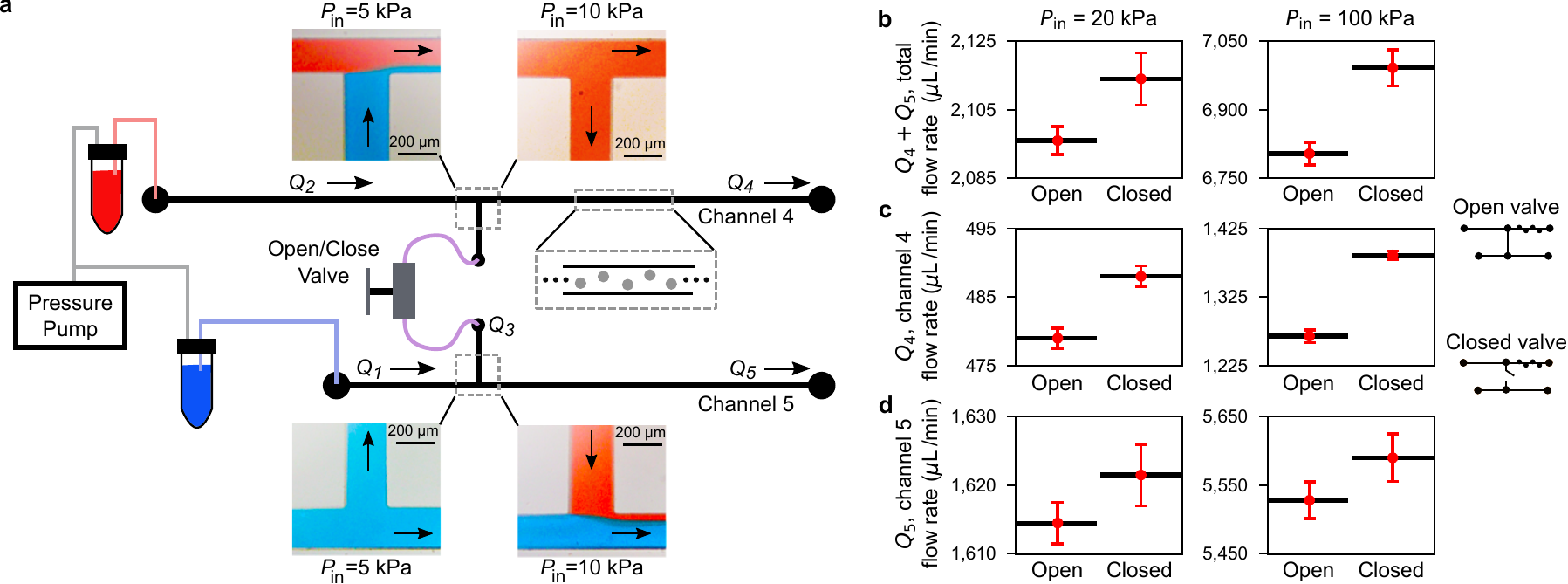}
  \caption{\textbf{Experimental observation of flow switch and Braess's paradox.} \textbf{a}, Experimental setup of the system presented in Fig.~\ref{fig1}a and flow tracking images at the junctions. An air-pressure pump is used to equally pressurize two vials containing red and blue dyed water, respectively, where each vial is connected to one of the system inlets. The linking channel is equipped with an open/close valve and channel 4 contains 20 obstacles. Images of the dyed flows through the junctions are shown for $P_{\mathrm{in}}$ below (5 kPa) and above (10 kPa) the flow switching pressure  $P_{\mathrm{in}}^*$, where the flow directions are indicated by the arrows.
\textbf{b},  Total flow rate ($Q_4+Q_5$) when the linking channel valve is ``Open" and ``Closed" for two different driving pressures above $P_{\mathrm{in}}^*$. 
\textbf{c}, \textbf{d},  Break down of the total flow rate into $Q_4$ (\textbf{c}) and $Q_5$ (\textbf{d}) for the two states of the valve. 
The plotted flow rates are averages derived from time series data, and the error bars indicate one standard deviation.
The observed increase in the total flow rate when the valve is closed is direct evidence of Braess's paradox.
}
\label{fig4}
\end{figure*}

\subsection*{Switching and Braess's paradox}
We incorporate the channel segment with obstacles characterized above into a network by considering the microfluidic system presented in Fig.~\ref{fig1}a.
We take the common static pressure  $P_{\mathrm{in}}$ at the inlets to be the controlled variable in the system.
The total flow rate through the network is now simply the sum of the flows at the outlets ($Q_4 + Q_5$).
In Fig.~\ref{fig3}, we present results for this system from direct simulations of the steady-state solutions of the Navier-Stokes equations.
As $P_{\mathrm{in}}$ is increased from zero, the  flow rate through the linking channel $Q_3$ is initially positive
before changing direction and becoming negative once a critical pressure, defined as $P^*_{\mathrm{in}}$, is reached (Fig.~\ref{fig3}). This flow switch results from the nonlinear change in pressure loss along the channel segment containing obstacles,
which causes a change in the sign of the pressure difference along the linking channel $\Delta P_{21}$  (approximately $P_2-P_1$) as the flow rate through the system increases with $P_{\mathrm{in}}$. 
We define $Q_{C}$ to be the total flow rate for the connected system configuration and $Q_{D}$ to be the total flow rate for the disconnected system configuration, where both are regarded as functions of $P_{\mathrm{in}}$.

Figure~\ref{fig3} shows $\Delta \mathcal{Q}\equiv Q_{C} - Q_{D}$ for a range of applied pressures $P_{\mathrm{in}}$.
Intuition may suggest that $\Delta \mathcal{Q}$ is positive for all values of $P_{\mathrm{in}}$ because
the linking channel in the disconnected system can be considered to have an infinite fluidic resistance, while for the connected system configuration the resistance of the linking channel is finite. Hence, reducing the resistance of any component of the system may seem to imply that the total flow rate should increase for fixed $P_{\mathrm{in}}$.  We observe, however, that $\Delta \mathcal{Q}$ becomes negative for $P_{\mathrm{in}}$ above the critical pressure that marks the flow switch, $P^*_{\mathrm{in}}$, meaning that an open linking channel between the parallel channels results in a {\it  lower} total flow rate. 
Figure~\ref{fig3} also shows that the flow rate through the channel segment with obstacles, $Q_4$, remains largely unchanged between the two configurations.
Therefore, the difference in the total flow rate exists primarily in the difference in $Q_5$, and $Q_3$ acts as a controlling variable of $Q_5$.

The observation of a lower total flow rate for the connected configuration compared to the disconnected configuration for fixed $P_{\mathrm{in}}$ is a manifestation of a fluid analog of Braess's paradox. Indeed, if we consider the disconnected system driven by an inlet pressure $P_{\mathrm{in}} > P^*_{\mathrm{in}}$, the addition of the linking channel can result in a significant {\it decrease} in the total steady-state  flow rate (as large as 10\% in our simulations). 
The value of the critical pressure $P^*_{\mathrm{in}}$ depends, of course, on the dimensions of the channels, but we find that the onset of  Braess's paradox and the flow switch always occur at the same pressure for the range of parameters investigated. 
We obtain similar results for Braess's paradox and flow switching
when instead the total pressure is controlled at the inlets (Supplementary Information, section \ref{S3.4}).
Our observation of Braess's paradox and flow switching also has the potential to lead to additional control features when existing microfluidic components are integrated into our system. For example, by incorporating an offset fluidic diode \cite{Adams2005} in the linking channel, the system can undergo negative (and positive) conductance transitions, where an \textit{increase} in $P_{\mathrm{in}}$ leads to an abrupt {\it decrease} in the total flow rate (Supplementary Information, section~\ref{S4}).

\subsection*{Experimental results}
We performed experiments to validate our predictions of flow switching and  Braess's paradox in a network with dimensions typical of microfluidics. A schematic of the experimental apparatus is presented in Fig.~\ref{fig4}a, where an open/close valve is used to implement the addition/removal of the linking channel (Methods). 
With the valve open, a flow switch is observed at a critical driving pressure $P_{\mathrm{in}}^*$ in the range of $5$--$10$ kPa, as demonstrated in Fig.~\ref{fig4}a by images of the flows through the channel junctions at the end points of this pressure range.
(The switching behavior has no reliance on the valve, as explicitly shown in Fig.~\ref{ExpSwitch}).

A confirmation of Braess's paradox in this system is shown in Fig.~\ref{fig4}b for driving pressures above $P_{\mathrm{in}}^*$, as observed in our simulations.
The measured total flow rate is higher when the linking channel valve is closed than when it is open, thus demonstrating the paradox, and the magnitude of the paradox is observed to be larger for higher driving pressures. A break down of how the flow rate changes in channel segments 4 and 5 individually is shown in Fig.~\ref{fig4}c,d. Closing the valve causes the flow rates through both channels to increase, which is in agreement with direct simulations and is yet another striking aspect of Braess's paradox in this system; it would be, at first, intuitive to expect that $Q_5$ would decrease when the in-flow from the linking channel is switched off. Time series of the flow rates measured as the linking channel is sequentially opened and closed further illustrate the transitions underlying the paradox (as shown in Fig.~\ref{ExpTimeSeries}).
 
In our experiments, the total pressure is controlled at the inlets and the experimental results
are in full qualitative agreement with simulations performed under the same pressure boundary conditions (Supplementary Information, section \ref{S3.4}).
This illustrates the robustness of the phenomenon, given that our simulations are in two dimensions and three-dimensional effects are expected to be significant in the experiments.
We note that different aspects of the paradox have been considered in fluid networks, but only 
for macroscopic (i.e., non-microfluidic) systems and while modeled by ad hoc flow equations \cite{calvertKeady,Penchina2009,Ayala2012}.
Analogs of the paradox have also been studied in several other areas, including electrical, mechanical, biological, and contemporary traffic networks \cite{Cohen1991a,Youn2008,Nicolaou2012,Pala2012,Motter2018}. 
These examples show that  Braess's paradox is a potentially general network phenomenon, which has remained unexplored in microfluidic networks.

\subsection*{Network model}
To characterize the microfluidic system in Fig.~\ref{fig1}a, we construct an analytic model that captures the flow properties observed in our simulations and experiments. The model consists of pressure-flow relations for each channel segment and, crucially, includes the most dominant term resulting from minor pressure losses at the channel junctions \cite{Crane1978,Khodaparast2014} (Methods). 
We model the contribution of the latter as
an additive term $K(Q_3/Q_1)f(Q_5)$ to the pressure-flow equation for channel segment $5$, where  the scaling factor $f$ and the coefficient $K$ are increasing functions for $P_{\mathrm{in}}\ge0$ such that $f(0)=K(0)=0$.
Several results are obtained from this model for $P_{\mathrm{in}}\,{>}\,0$, as assumed throughout. 
First, if $\beta=0$ (i.e., the quadratic term is zero in equation~(\ref{eq2})) when the static pressure is controlled or the dynamic pressure is negligible, then flow switching does not occur, in agreement with direct simulations (Supplementary Information, section~\ref{S3.2}).
Second, when $\beta > 0$, a 
steady-state solution can be found 
satisfying $Q_{\mathrm{3}} = 0$
provided that the following geometric condition is satisfied: 
\begin{equation} \label{eq3}
L_{1} < \frac{12 L_{2} L_{5}}{\alpha w^2 L_{4}} = L^*.
\end{equation} 
This solution identifies the critical pressure $P^*_{\mathrm{in}}$. 
Third, for flow rates in the linking channel, the model predicts that a variation $\delta Q_{3}$ is negatively related to a variation $\delta P_{\mathrm{in}}$ around $P^*_{\mathrm{in}}$.
This indicates that $P_{\mathrm{in}}$  
above (below) $P^*_{\mathrm{in}}$ results in a negative (positive) flow rate through the linking channel.
The first result implies that, in our experiments, 
the Forchheimer effect is necessary to achieve a flow switch.
The second and third results, which hold even for when dynamic pressure is non-negligible, show that this model captures the flow switching behavior observed in the simulations and experiments. 
Importantly, we validate the flow switching condition in equation (\ref{eq3}) 
by demonstrating quantitative agreement between the model and simulations both when the static and when the total pressure is controlled  (Supplementary Information, section~\ref{S3.2}).

The model also predicts Braess's paradox as observed in our experiments and  simulations. Specifically, under the condition that equation~\eqref{eq3} is satisfied and dynamic pressure is small (or static pressure is controlled),
the model predicts the paradox to occur for $\delta P_{\mathrm{in}}>0$   if and only~if
 \begin{equation}
 K'(0) \beta f\bigg(\frac{a}{\beta}\bigg) > c,
 \label{eq4}
 \end{equation}
 where $a$ and $c$ are positive parameters and prime denotes derivative. 
If total pressure is controlled and dynamic pressure terms are included, the paradox is also predicted for $\delta P_{\mathrm{in}}>0$ provided that a relation similar to equation~(\ref{eq4}) is satisfied  (details for both cases are presented  in Supplementary Information, section~S2). 
The dependence of condition (\ref{eq4}) on $\beta$ and $K'(0)$ underlines the crucial role of nonlinearity and minor losses in 
 giving rise to Braess's paradox in our experiments, and shows in particular that minor losses have to be sufficiently large. 
 Indeed, if the effect of minor losses is neglected, a 
 manifestation
of Braess's paradox  is 
still predicted to occur, but with much smaller magnitude and only for  $\delta P_{\mathrm{in}}<0$, which is inconsistent with our simulations and experiments (Supplementary Information, section~S2.3).

 The result in equation~\eqref{eq4} also highlights a fundamental difference between microfluidic and electronic circuits, 
 namely that minor losses (i.e., significant energy losses associated with interactions between circuit components) do not 
 have direct analogs in common electronics.
Given the central role played by such losses in equation~\eqref{eq4}, we posit that this difference might be the reason why no equivalent of 
 the Braess paradox effect we present has been observed in electronic networks, even though 
 aspects of it have \cite{Cohen1991a}. 
 We further investigated the impact of interactions between channel segments by varying the junction angles to show that the paradox can be further enhanced by manipulating the minor losses (Supplementary Information, section~\ref{S3.3}).

\begin{figure*}[htb] 
\centering 
\includegraphics[width=0.99\textwidth]{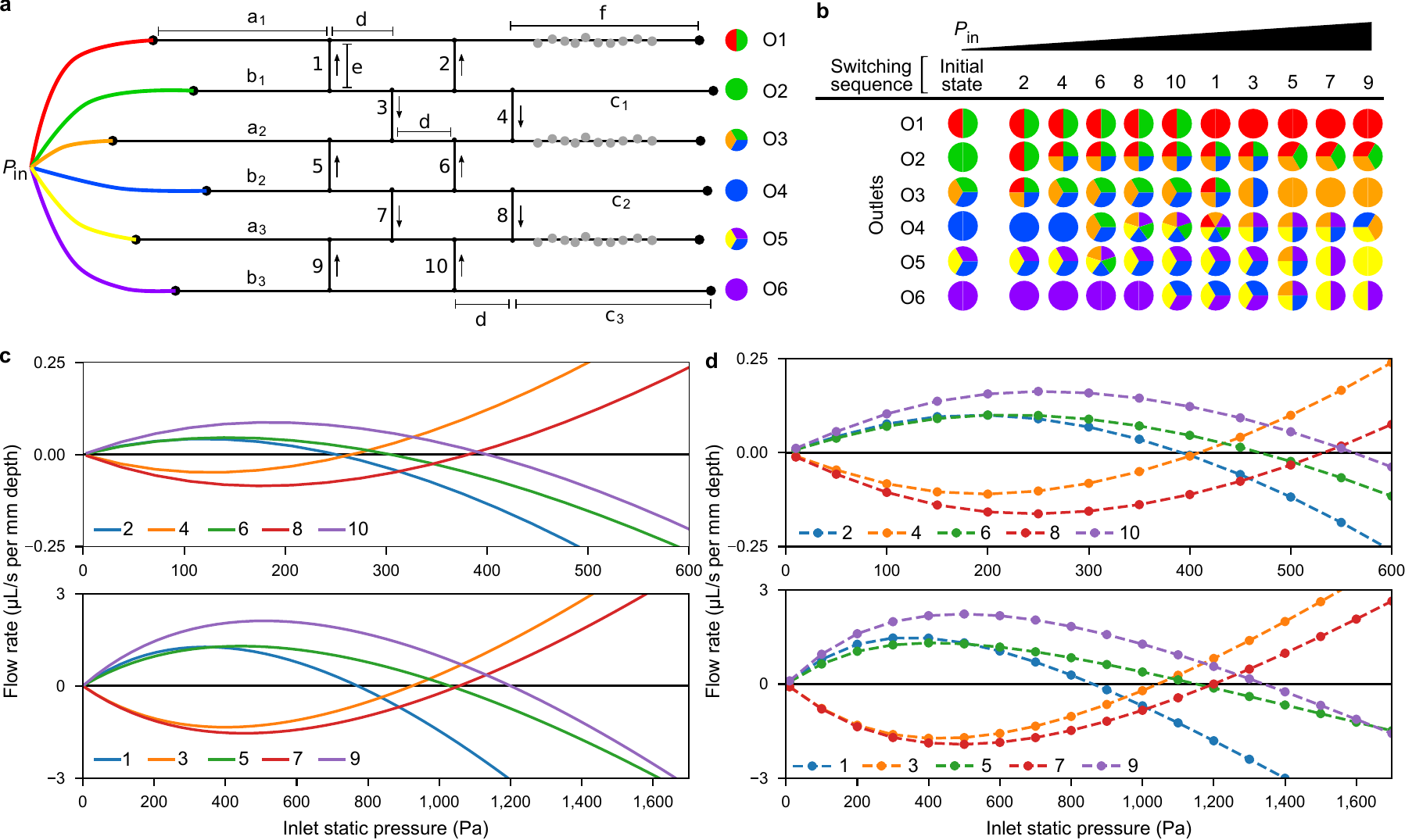}
\caption{\textbf{Flow patterns in a multiswitch network.}
\textbf{a}, Schematic of ten-switch network. Fluids of different colors are driven to each inlet by a common static pressure source, $P_{\mathrm{in}}$. The outlets are labeled by O1--O6 and the linking channels by 1--10. The arrows indicate the flow direction through each linking channel and multicolored circles schematically indicate the fluid composition at each outlet for an initially low $P_{\mathrm{in}}$. The  segment lengths are denoted by $a_i, b_i, c_i, d, e,$ and $f$, where the segments with obstacles are marked with gray circles, and a common length is assumed for all linking channels.
\textbf{b}, Patterns of outlet flows for the network programmed with a chosen switching sequence as $P_{\mathrm{in}}$ is increased. 
Each column of colored circles denotes the outlet flows after the corresponding flow switch occurs, where  mixing between different colored fluids is assumed to occur when passing through the same channel segment.
\textbf{c,d}, Model predictions (\textbf{c}) and simulation results of the Navier-Stokes equations (\textbf{d}) for the flow rate through each linking channel
for a network designed to exhibit the switching sequence in \textbf{b}.
The flow rates are labeled according to the channels in \textbf{a} and are divided into two sets (top and bottom panels) for clarity. Positive flow rates correspond to flow in the upward direction in  \textbf{a}, and each flow switch occurs when the corresponding curve crosses the horizontal axis. The segment dimensions that give rise to the particular switching order in \textbf{b}-\textbf{d} are reported in Table~\ref{multiswitchTable}. 
All twenty-one possible outlet flow color combinations are realized between the switching sequence presented here and those in Fig.~\ref{suppSeq}.
}
\label{fig6}
\end{figure*}

 \subsection*{Networks with multiple programmed switches}
The system considered thus far can be generalized to create larger microfluidic networks with multiple flow switches. That is, networks with multiple disjoint channel segments in which the flow initially in one direction can be individually ``switched" to move in the opposite direction through the manipulation of one driving pressure alone. 
In our design, the linking channel plays the role of a switch (and can be referred to as such).
Figure~\ref{fig1}b shows the multiswitch generalization of the network in Fig.~\ref{fig1}a, which incorporates multiple linking channels and a subset of channel segments with obstacles. We experimentally demonstrate an instantiation of a six-switch network that exhibits flow switching in  all linking channels (as presented in Supplementary Information, section~\ref{S6.2}). Multiswitch networks can be 
designed by extending the network model presented above.

One such network with ten linking channels is presented in Fig.~\ref{fig6}a.
By marking each inlet flow with a different color, we show that a variety of patterns can form in the outlet flows (colored circles in Fig.~\ref{fig6}). The specific pattern at an outlet depends on the order in which the flow switches occur as  $P_{\mathrm{in}}$ is varied.
The network model for larger systems is constructed by combining pressure-flow relations for each channel segment with flow rate conservation equations for each junction. 
Using this model, 
we can design 
a network for which each flow switch occurs near a target value of $P_{\mathrm{in}}$ by optimizing the dimensions of the channel segments  (Methods).

As illustrated in Fig.~\ref{fig6},
a set of eleven different internal flow states and seventeen unique color combinations at the outlets are possible 
for the switching sequence realized  in Fig.~\ref{fig6}b.
Figure~\ref{fig6}c,d shows the agreement between the model predictions of these flow states 
and results from direct simulations of the Navier-Stokes equations.
This variety of states (and output patterns) is achieved with only three channel segments containing obstacles and is parameterized by 
a single  control variable---the driving pressure $P_{\mathrm{in}}$.
 Moreover, the switching is implemented solely through the working fluid, which differs from existing approaches that rely on flexible valves and additional control flows \cite{Leslie2009}.
 Thus, multiswitch networks exhibit several properties exploitable in the design of new controllable microfluidic systems.

More generally, for a multiswitch network with $n_c$ horizontal channels interconnected by $n_l$ linking channels, the number of possible internal flow states is $n_l +1$ if each linking channel exhibits a flow switch. In addition, the possible number of unique color combinations in the outlet flows is $n_c(n_c+1)/2$ if each inlet flow is marked with a different color.
All color combinations can be realized over the set of all switching sequences, provided that there exists flow paths allowing mixing of every set of $k$ adjacent colors for $k$ ranging from $1$ to $n_c$. 
The myriad of states possible in such multiswitch networks underlies their ability to process inputs into multiple outputs and thus to support various applications,
including implementing different mixing orders of chemical reagents and devising schemes for the parallel generation of mixtures with tunable concentrations.

\subsection*{Conclusions and outlook}
The flow switch, conductance transitions, and Braess paradox established in this study are all emergent behaviors of common origin resulting from nonlinearity and interactions between different parts of the system.  
The nonlinearity is directly determined by fluid inertia effects, 
which can be enhanced and manipulated through the placement of obstacles and has the advantage of not being reliant on flexible components, fluid compressibility, or dedicated control flows. 
The onset of Braess's paradox is marked by the flow switching pressure, above which the increased resistance of the nonlinear channel causes the flow to be routed in the negative direction through the linking channel. 
When constrained by a diode, the switch in flow direction also enables negative conductance transitions.
Our results demonstrate an approach for routing and switching in microfluidic networks 
through control mechanisms that are coded into the network structure,
thus responding to the call for design strategies that allow diverse microfluidic systems to be assembled from a small set of core components~ \cite{Stone2009,Bhargava2014}.

Here, we considered the scenario in which the inlets and the outlets are (separately) held at the same pressure, rendering the network a two-terminal system in all cases, since this is the most stringent scenario for flow manipulation. 
If a multi-terminal system is configured, by allowing the pressures at each of the inlets (and/or outlets) to be varied independently, then the effects we presented may be further enhanced. 
Finally, while we focused on boundary conditions in which the inlet pressures are controlled, it would be natural to explore in future research the scenario in which the controlled variables are the inlet flow rates. We anticipate, for example, that the negative conductance transitions are then converted into pressure amplification (pressure release)  transitions in which the inlet-outlet pressure difference increases (decreases) abruptly at the transition point.
Accordingly, the Braess paradox is also expected to take a complementary form in which closing the linking channel causes the inlet-outlet pressure difference to drop. Incidentally, it is this complementary form of Braess's paradox that has been previously established for electric circuits \cite{Cohen1991a}, thus suggesting an additional correspondence between electronic and microfluidic circuits.

\subsection*{Acknowledgments}
This research was supported by the National Science Foundation under Grants Nos.\ PHY-1001198 and CHE-1465013, the Simons Foundation through Award No.\ 342906, and a Northwestern University Presidential Fellowship.  

\subsection*{Methods}
\vspace{-4mm}
\textbf{Navier-Stokes simulations.}
The numerical simulations were performed using OpenFOAM-version 4.1 \cite{foam}. We used meshes with an average cell area ranging from $10\,\mu$m$^2$ to $340\,\mu$m$^2$, where the finer meshing was applied near the obstacles. All meshes were generated using Gmsh \cite{Geuzaine2009}. The two-dimensional solutions were found using the simpleFoam solver within OpenFOAM, employing second-order numerical schemes, where a fixed static pressure of zero was set for the boundary conditions at the outlets. At the inlets, the static (total) pressure was fixed for the static (total) pressure controlled cases. For simulations of the multiswitch network in Fig.~\ref{fig6}, the same geometry and dimensions were used as for the model predictions, provided in Table~S1, and equal driving pressures were applied at each of the six inlets.

\noindent
\textbf{Reynolds numbers.} The characteristic length scale used in defining the Reynolds number of the flow is the hydraulic diameter of the channels, defined as $4 A/P$, where $A$ is the area and $P$ is the perimeter of the channel cross section (common to all segments). The hydraulic diameter in two and three dimensions is $2w$ and $2wh/(w+h)$, respectively, where $h$ is the height of the channels in the three-dimensional case. The characteristic velocity used in two and three dimension is $Q/w$ and $Q/wh$, respectively. Therefore, we define $Re=2\rho Q/\mu$ for our simulations in two dimensions and $Re=2\rho Q/\mu(w+h) $ for our experiments  in three dimensions.
    The undeclared ranges of $Re$ for the channel segment with obstacles considered in the presented data are: $21$--$385$ (Fig.~\ref{fig3}), $12$--$121$ (Fig.~\ref{fig4}),  $1$--$220$ (Fig.~\ref{fig6}), $1$--$380$ (Fig.~\ref{paradox_noML}), $4$--$111$ (Fig.~\ref{figS4}), $40$--$385$ (Fig.~\ref{paradox_slanted}), $20$--$400$ (Fig.~\ref{paradoxTP1}), $2$--$10$ (Fig.~\ref{ExpSwitch}b), $75$--$85$ (Fig.~\ref{ExpSwitch}c), $76$--$89$ (Fig.~\ref{ExpTimeSeries}),  $10$--$20$ (Fig.~\ref{ExpMultiSwitch}b), and $110$--$120$ (Fig.~\ref{ExpMultiSwitch}c).

\noindent
\textbf{Pressure boundary conditions.} 
We consider two different boundary conditions for the driving pressure $P_{\mathrm{in}}$ at the system inlets. Under one condition, total pressure is controlled and the  inlets open directly into a high-pressure reservoir. Under the other condition, static pressure is controlled and the inlets are connected to the reservoir by pressure regulators. Total pressure is the sum of static pressure and dynamic pressure, where dynamic pressure is defined as $\frac{1}{2}\rho v^2$ for a fluid with density $\rho$ and velocity $v$. The distinction between these boundary conditions is often neglected in the microfluidics literature when the Reynolds number is less than one \cite{Oh2012}, but it can become important for larger Reynolds numbers
(even though the flow remains laminar) \cite{Zeitoun2014}.

\noindent
\textbf{Pressure-flow relations for microfluidic channels.} 
We use equation (\ref{eq1}) to describe the pressure-flow relation for straight, obstacle-free channels, which is derived directly from the Navier-Stokes equations by assuming plane Poiseuille flow through a two-dimensional channel.
To describe the nonlinear pressure-flow relation observed for the channel with obstacles we refer to the Forchheimer equation:
$
-\Delta P=\alpha\mu L V + \beta\rho L V^2
$,
where $V$ is the average fluid velocity. 
In two dimensions, $V = Q/w = \mu Re/2\rho w$ and, thus, the Forchheimer equation can be written in the form of equation~(\ref{eq2}).
In agreement with 
equation~(\ref{eq2}), we find an excellent linear fit between $-\Delta P/Re$ and $Re$ for a channel with ten obstacles, and we validate the fit by predicting flows through the same channel for a fluid with a different viscosity (Supplementary Information, section~\ref{S3.1} and  Fig.~\ref{fig5SI}b).
We observe no unsteady flow through the channel with obstacles due to vortex shedding for $Re$ of up to $400$, as expected for systems with highly confined obstacles \cite{Zovatto2001}, which permits the use of the steady-state relation in equation~(\ref{eq2}) over the range of $Re$ considered here. 
We experimentally verify the source of nonlinearity in PDMS channels with obstacles, which were designed to have approximately square cross-sections to minimize deformation (which could lead to other forms of nonlinearity \cite{Gervais2006,Christov2018}).
Through additional experiments, we confirmed 
that pressure-flow relations similar to those in Fig.~\ref{fig2}e,f
hold for channels constructed from materials with both higher rigidity (SU-8 photoresist) and lower rigidity (Flexdym) than the PDMS (Supplementary Information, section~\ref{S5} and Fig.~\ref{ExpFlex3d}).
We note that porous-like structures have been previously used both to study non-inertial effects in microfluidics, such as droplet formation \cite{Amstad2014} and viscous fingering \cite{Haudin2016}, and to study inertial effects in larger systems \cite{Zhao2016}. In our system, inertial effects arise at the microfluidic scale even for a much smaller number of obstacles than the typical number in porous-like materials.

\noindent
\textbf{Network flow model construction.}
The analytic model used to describe the system in Fig.~\ref{fig1}a
is constructed as follows:
(i) we consider the  pressure at the inlets $P_{\mathrm{in}}$ to be in the vicinity of $P^*_{\mathrm{in}}$;
(ii) we approximate the pressure-flow relation  through the linking channel as $Q_{3} = \kappa (\gamma P_1 - P_2)$, where 
$\kappa$ is the channel conductivity
and $\gamma$
is a free parameter allowing for an effective pressure difference;
(iii) the flow equation for each other channel segment without obstacles is written as in equation~(\ref{eq1}), where $-\Delta P$ is the pressure drop along the segment and $L$ is the segment length; 
(iv) for the channel segment with obstacles, we take the flow equation to be in the form of equation~(\ref{eq2}) (with $Re$ expressed as $2\rho Q/\mu$);
(v) we include the most dominant term resulting from minor pressure losses at the channel junctions.
Therefore, the model consists of five pressure-flow relations, in addition to  
two flow conservation equations at the junctions: $Q_{3} + Q_2 - Q_4 =0$ and $Q_{3} + Q_5 - Q_1 =0$.
When the static pressure is controlled at the inlets, the only nonlinearity that exists in the model comes from the Forchheimer term due to the presence of obstacles and the minor loss term. 
The model can also be adapted for when total pressure is controlled by taking the static pressure at each inlet to be $P_{\mathrm{in}} -  k\rho Q^2/2 w^2$,  where $P_{\mathrm{in}}$ now denotes total pressure and the coefficient $k$ is a constant of order unity that only depends on the shape of the inlet velocity profile ($k\approx1$ for a uniform velocity profile at the inlet, as considered here).
However, the dynamic pressure  term $\rho Q^2/2w^2$ is often negligible in real microfluidic systems because of the high pressures needed to drive fluid though the channels. Indeed, in our experiments, 
the dynamic pressure near $P_{\mathrm{in}}^*$ was smaller than the static pressure by two orders of magnitude and smaller than the pressure loss due to the Forchheimer effect by one order of magnitude. This can also be seen in Fig.~\ref{fig2}f, where a constant relation between $Re$ and $\Delta P/Re$ is measured.
Details of the model  are presented in Supplementary Information, section~S1.

\noindent
\textbf{Designing multiswitch networks.}
For a network with multiple switches and a  
given set of channel dimensions, the value of $P_{\mathrm{in}}$ for which a specific flow switch occurs can be determined through the addition of a constraint to the model that enforces the flow through the corresponding linking channel to be zero.
Then, the dimensions of a chosen subset of channel segments may be iteratively varied through an optimization procedure in order to design a network for which each flow switch occurs near a target value of $P_{\mathrm{in}}$.
Depending on which dimensions are allowed to be adjusted,
the desired relative order of the switches can be achieved exactly, and the final set of switching pressures can be very close to the target ones (often $<5\%$ difference),
where the former is expected to be more important in applications.
Further details on the design of multiswitch networks are presented in Supplementary Information, section~\ref{S6.1}.

\noindent
\textbf{PDMS channel fabrication.}
The flow channels were assembled by sealing a patterned PDMS 
chip against a glass slide. The PDMS chip was made by pouring a mixture of PDMS oligomer and cross-linking curing agent (Sylgard 184) at a weight ratio of 10:1 into a mold after being degassed under vacuum. The mixture was cured at $74\,\degree$C for $1$ h and then peeled off from the mold to yield the microchannel design. The dimensions of the channels in Figs.~\ref{fig2} and \ref{fig4} were $200\,\mu$m (width) $\times$ $185\,\mu$m (height), and the diameter of the obstacles was $97\,\mu$m. 
After punching the holes for inlet and outlet connections, the PDMS chip was thermally aged at $200\,\degree$C for $12$ h to reduce pressure-induced deformation \cite{Kim2014}, yielding a chip with a Young's modulus of approximately 3 MPa \cite{Johnston_2014}. Both the PDMS chip and the glass substrate were cleaned with isopropanol and treated by plasma for $90$ s before bringing them into contact. Once the PDMS chip was sealed against the glass slide, the device was placed in an oven for $30$ min at $74\,\degree$C to improve bonding quality.

The mold used was a silicon wafer containing microchannel patterns created by soft photolithography using a negative photoresist \cite{Martin2000,Duffy1998}. A 4-inch silicon wafer (test grade, University Wafer, Boston, MA) was cleaned with acetone and isopropanol and dried with nitrogen gas. The wafer was then coated with SU-8 50 negative photoresist (MicroChem Corp., Newton, MA) on a spin coater (Laurell Technologies Corp., North Wales, PA) operating at $600$ rpm for $30$ s. After a pre-exposure bake at $65\,\degree$C and subsequently at $95\,\degree$C, each for $60$ min, the coated wafer was exposed to UV light (Autoflood 1000, Optical Associates, Milpitas, CA) through a negative transparent photomask that contained the desired channel design. Following a 3.5 min post-exposure bake at $95\,\degree$C, the wafer was developed in SU-8 developer (MicroChem Corp., Newton, MA) for $60$ min to obtain the pattern.

\noindent
\textbf{Flexdym channel fabrication.}
Flexdym (Blackholelab Inc., Paris) is a thermoplastic elastomer (Young's modulus of 1.18 MPa) with a rapid and easy molding process for microfluidic devices \cite{Lachaux2017}. After fabrication of the silicon wafer mold containing the channel designs, a sheet of Flexdym (6 cm $\times$ 4 cm) was placed directly above the mold with another sheet of unpatterned PDMS (about 1 mm thick) placed above the Flexdym for protection. The whole set was then placed on a heat press between two Teflon sheets. The plate on the heat press was heated to 175$\degree$C before starting to mold the Flexdym. Once the target temperature was reached, the lever on the heat plate was locked down with a timer set for 5 min. After the process was finished, the lever was released and the Flexdym sheet was inspected visually to make sure that no bubbles were trapped around the channel. The chip was allowed to cool down for 5 min before unfolding the layers. 
The Flexdym was permanently sealed with a glass slide by following the same sealing procedure used for the PDMS channels.
The dimensions of the cross-section of the  channels were $201\,\mu$m (width) $\times$ $166\,\mu$m (height), and the diameter of the obstacles was $99\,\mu$m.

\noindent
\textbf{SU-8 photoresist channel fabrication.}
To make microfluidic channels directly from SU-8 photoresist, an inverse mask was designed and printed on transparency. The desired channel was printed on the inverse mask in black with transparent dots marking the obstacles, and the rest of the mask was left transparent. The same procedure to make the silicon wafer master as described in ``PDMS channel fabrication" was followed to fabricate the channels on glass slides. The chip was then sealed by 3M VHB tape to another glass slide with holes for connections.
The dimensions of the cross-section of the  channels were $209\,\mu$m (width) $\times$ $196\,\mu$m (height), and the diameter of the obstacles was $90\,\mu$m.
 The Young's modulus of SU-8 photoresist is 2 GPa  (from table of properties for SU-8 permanent photoresists, MicroChem Corp., Newton, MA).

\noindent
\textbf{Flow rate measurement.}
Experimental measurements in Figs.~\ref{fig2} and \ref{fig4} were made with the system shown in Fig.~\ref{fig4}a. When measuring the relation between pressure and flow rate, the linking channel valve was closed to allow separate measurement of the channel with and the channel without obstacles. Deionized (DI) water was pumped through each channel and a pressure scan from $0$ to $100$ kPa was performed using an Elveflow OB1 pressure controller. The flow rate was measured by an Elveflow MFS5 flow sensor ($0.2$ - $5$ mL/min). To verify Braess's paradox, the same instruments were used and the pressure was set constant while recording the flow rate at each outlet. Red (3 g/L, FD\&C Red \#40, Flavors \& Colors) and blue (1.5 g/L, FD\&C Blue \#1, Flavors \& Colors)
dyes were added into DI water to demonstrate the switching behavior. The concentrations of the dyes were adjusted for similar flow rate under the same pressure.
The flow rate measurements in Fig.~\ref{ExpFlex3d} were performed using isolated channels constructed from Flexdym and SU-8 photoresist, respectively.

\noindent
\textbf{Fluorescence imaging.}
Fluorescent polyethylene microspheres ($10$-$20\,\mu$m) were suspended in Tween 80 solution (Cospheric LLC, Santa Barbara, CA)  
and pumped through a single microfluidic channel with obstacles by an Elveflow OB1 pressure controller. Two different pressures were applied,  $3$ kPa and $100$ kPa, to demonstrate different flow profiles around the obstacles. Fluorescence images were captured with an Olympus BX51 microscope equipped with a NIBA filter through an Infinity 3 CCD camera.

\noindent 
\textbf{Measured flow rate data and statistics.} Savitsky-Golay filtering was applied to all flow rate data collected through experiments, using a window length of $11$ data points and a second-order polynomial.
For each of the fixed pressures presented in Fig.~\ref{fig4}b-d, a $60$ s time series of flow rate data was collected at each of the outlets with a sampling rate of $10$ Hz. Over the $60$ s interval, the linking channel valve was sequentially opened/closed every $15$ s. For each time series, the $15$ s intervals in which the valve was open (closed) were averaged to create a single 15 s time series for each outlet. The total flow rate ($Q_4 + Q_5$) was calculated when the valve is open and closed, respectively, by summing the 15 s time series for the two outlets point-by-point. The statistics presented in Fig.~\ref{fig4} are the average and standard deviation of the resulting series.
For Fig.~\ref{ExpTimeSeries}, the flow rate at each of the two outlets was measured experimentally at a sampling rate of 100 Hz over a $180\,$s interval, during which the linking channel was sequentially opened/closed every $30$ s. The total flow rate in Fig.~\ref{ExpTimeSeries}c was calculated by summing, point-by-point, the data in Fig.~\ref{ExpTimeSeries}a and b.

\noindent
\textbf{Parameters in simulations and experiments.} 
In the simulations, we set $\rho =10^{3}\,$kg/m$^3$, $\mu = 10^{-3}\,$Pa$\cdot$s, $\nu = \mu/\rho =10^{-6}\,$m$^2$/s, $w=500\,\mu$m for the width of all channels, and $r=100\,\mu$m for the radius of all obstacles, unless otherwise noted. 
In all experiments, DI water was used as the working fluid.
The other undeclared dimensions were as follows. In Fig.~\ref{fig2}a,b, the length of the (partially shown) channel was $1.25\,$cm. 
In Fig.~\ref{fig2}c-e, the channel length was $4.3\,$cm, and in Fig.~\ref{fig2}f  the channel length was $2.0\,$cm (see PDMS channel fabrication for the remaining dimensions). 
In Fig.~\ref{fig3}, $L_1 =0.17\,$cm, $L_2 =1.0\,$cm, $L_3 =0.1\,$cm, $L_4 =1.25\,$cm, and $L_5 =1.0\,$cm. 
In Fig.~\ref{fig4}, $L_1 =0.6\,$cm, $L_2 =2.9\,$cm, $L_4 =1.4\,$cm, and $L_5 =1.4\,$cm.  
For the linking channel, the switch valve was connected to the two parallel channels through $15\,$cm of round tubing and $0.7\,$cm of microchannel on each side. Each inlet was connected to the pressurized vials through $62\,$cm of tubing, and each outlet was attached to $50\,$cm of tubing. The inner diameter of all tubing was $0.79\,$mm.

\let\oldaddcontentsline\addcontentsline
\renewcommand{\addcontentsline}[3]{}

\let\addcontentsline\oldaddcontentsline

\newpage
\onecolumngrid
\clearpage

\setcounter{page}{1}
\renewcommand{\theequation}{S\arabic{equation}}
\setcounter{equation}{0}
\renewcommand{\figurename}{Supplementary Fig.}
\setcounter{figure}{0}
\renewcommand{\tablename}{Supplementary Table}
\setcounter{table}{0}
\renewcommand{\vec}{\mathbf}

\setcounter{section}{0}
\renewcommand{\thesection}{S\arabic{section}}

\begin{center}
{\bf\sc\LARGE Supplementary Information}

\bigskip
{\it Braess's paradox and programmable behaviour in microfluidic networks
}
\end{center}

\tableofcontents

\section{Model of fluid system}\label{S1}

\noindent
A network schematic of the system in Fig.~\ref{fig1}a of the main text is shown in Supplementary Fig.~\ref{schem}. The inlets are driven by a pressure $P_{\mathrm{in}}$ relative to a constant static pressure $P_{\mathrm{out}}$ at the outlets, which is taken to be zero without loss of generality. The system includes two internal channel junctions, corresponding to pressures $P_1$ and $P_2$.
All channels have width $w$, and the length of each segment is denoted by $L$. The pressure loss along the linking channel is considered in section~\ref{S1.1} (minor pressure losses due to the internal junctions are considered in section~\ref{S2}). For all other obstacle-free channels, the pressure loss will be approximated by the Poiseuille law in two-dimensions:
\begin{equation}
\frac{- \Delta P}{L} = \frac{12 \mu }{w^3} Q,
\label{Poiseuille}
\end{equation}
where $\mu$ is the dynamic viscosity.
For the channel segment with obstacles, we make use of the Forchheimer equation:
\begin{equation}
\frac{P_2-P_{\mathrm{out}}}{L_{4}} = \frac{\alpha \mu}{w} Q_{4} + \frac{\beta\rho}{w^2} Q_{4}^2,
\end{equation}
where $\rho$ is the fluid density, and $\alpha$ and $\beta$ are parameters determined using direct numerical simulations.

\begin{figure}[htb] 
\centering 
\includegraphics[width=0.8\textwidth]{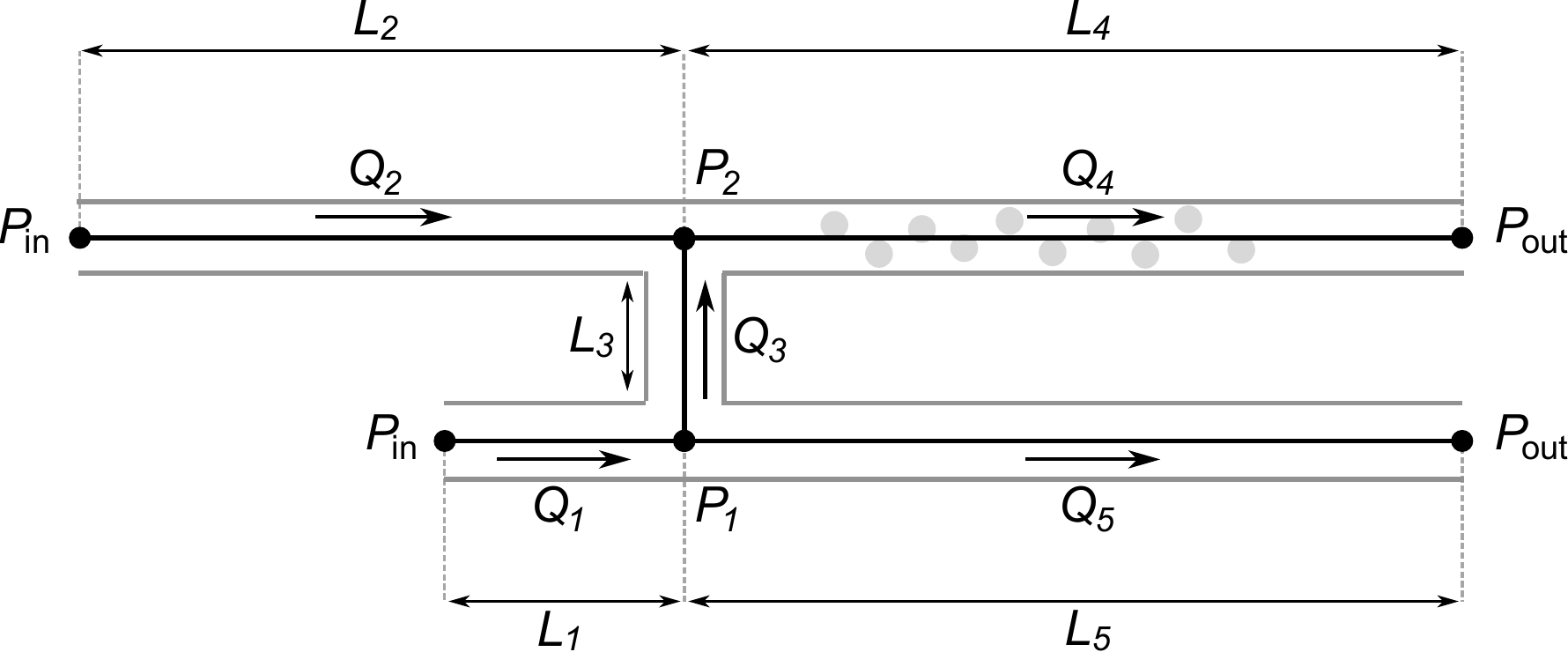}
\caption{\textbf{Network representation of the  system in Fig.~\ref{fig1}a}. The flow rate through each segment is indicated by $Q$, the channel lengths by $L$, the pressures by $P$, and the positive flow directions by arrows.} 
\label{schem}
\end{figure}

In order to simplify the equations used to describe this system, we first define the non-dimensional pressure $\overline{P}$, flow rate
$\overline{Q}$, and channel length $\overline{L}$ as 
 \begin{equation}
 \overline{P} = \frac{P}{\rho \nu^2/w^2}, \, \overline{Q} = \frac{Q}{\nu}, \, \overline{L} = 12 \frac{L}{w},
 \end{equation}
where $\nu=\mu/\rho$ is the kinematic viscosity. Here, we used that $Q$ has the dimensions of $\nu$ in two-dimensional flows.
The pressure loss equations then take the simpler form
\begin{equation}
 -\Delta \overline{P} =  \overline{L} \ \overline{Q}
 \end{equation}
 in obstacle-free channels, and the form
 \begin{equation}
\overline{P}_2-\overline{P}_{\mathrm{out}} = \frac{1}{12}\overline{L}_{4} ( \overline{\alpha}  \overline{Q}_{4} + 
\overline{\beta} \ \overline{Q}_{4}^2)
\label{eq:5}
 \end{equation}
for $\overline{\alpha} = \alpha w^2$ and $\overline{\beta} = \beta w$ in the channel with obstacles. 

Going forward,  all variables are considered to be non-dimensional, unless stated otherwise.  
The bars over the non-dimensional variables will be omitted for brevity. We use index $i$ to indicate $1, 2, 3, 4, 5$ and, in the case of $P$, to also indicate {\it in} and {\it out}. Moreover, we focus on solutions for $P_{\mathrm{in}}>0$ (for both the static and the total pressure at the inlets).
 
\subsection{Flow switching under controlled static pressure}\label{S1.1}

\noindent
When the static pressure is the controlled variable at the inlets, all $P_i$ are taken to be static pressure values. The non-dimensional model is composed of the pressure loss equations for the five channel segments together with the flow rate conservation equations at the two internal junctions. The resulting set of equations reads 
 \begin{eqnarray}
 P_{\mathrm{in}} - P_1 -  L_{1} Q_{1} &=& 0, \label{model1}\\
 P_{\mathrm{in}} - P_2 -  L_{2} Q_{2} &=& 0,\label{model2}\\
 P_1 - P_{\mathrm{out}} - L_{5} Q_{5} &=& 0, \label{model3}\\
 P_2-  P_{\mathrm{out}}  - \frac{1}{12}  L_{4} (\alpha   Q_{4} +  \beta {{Q}_{4}^2}) &=& 0, \label{model5}\\
 Q_{\mathrm{3}} - \kappa (\gamma P_1 - P_2) &=& 0,\label{model6}\\
 Q_{\mathrm{3}} + Q_{2} -  Q_{4} &=& 0, \label{model7}\\
 Q_{\mathrm{3}} + Q_{5} -  Q_{1} &=& 0, \label{model8}
 \end{eqnarray}
where the parameter $\kappa$ in equation~(\ref{model6}) is the hydraulic conductivity of the linking channel and the parameter $\gamma$ allows for an effective pressure difference that governs $Q_3$. 
Given the comparatively short length and wide width of the linking channel,  $\kappa$ may deviate from $1/L_3$ and will generally depend on the dimensions of the linking channel. Similarly, $\gamma$ may deviate from 1 because 
the magnitude and direction of the flow through the linking channel may be offset from those predicted by the point pressures ($P_1$ and $P_2$) due to the finite size of the junctions.
As shown in Supplementary Fig.~\ref{schem}, we consider the static pressure at the inlets $P_{\mathrm{in}}$ to be equal (and assumed to be tunable). Similarly, the outlets are connected to a common low-pressure reservoir ($P_{\mathrm{out}} = 0$, as noted above).   

By setting $\vec X = (P_1,P_2,Q_{1} ,  Q_{2},Q_{3},  Q_{4} , Q_{5} )^T$, equations~(\ref{model1})-(\ref{model8}) take the form
\begin{equation}
\vec G(\vec X, P_{\mathrm{in}}) = \vec 0,
\label{G}
\end{equation}
where $\vec{G} : \vec{R}^7 \times \vec{R} \rightarrow \vec{R}^7$. 
As stated in the main text, by adding the constraint $\gamma P_1 = P_2$ (equivalent to setting $Q_{\mathrm{3}} = 0$) to the system $\vec{G}$, we can determine the critical value $P_{\mathrm{in}}= P^*_{\mathrm{in}}$ at which the flow
switch through the linking channel occurs. In Supplementary Fig.~\ref{figS4}, we use $\gamma$ to fit the model prediction of $P_{\mathrm{in}}^*$ to simulation results. For all discussion that follows, we take $\gamma = 1$ to simplify the analysis. In solving for the flow switching point, we note that zero for all variables is always a solution, but this solution is trivial in that it corresponds to no flows through the system. 
A solution for $P^*_{\mathrm{in}}>0$, and thus $X_j > 0 $ for all $j\neq 5$ ($X_5=Q_{3} = 0$), can be found only if
\begin{equation}
L_{1} < \frac{12 L_{2} L_{5}}{\alpha L_{4}}.
\label{L1A}
\end{equation}
Equation (\ref{L1A}) is the non-dimensional counterpart  to equation~(3) in the main text and provides a geometric restriction on the system that must be satisfied in order to observe a switch in the flow direction through the linking channel, for a strictly positive driving pressure $P_{\mathrm{in}}$. Otherwise, the flow rate $Q_3$ is  negative  for all $P_{\mathrm{in}} > 0$. When the condition in equation~\eqref{L1A} is satisfied, the expression for $P_{\mathrm{in}}^*$ takes the form 
\begin{equation}\label{p1star}
P_{\mathrm{in}}^* = \frac{F_1(\alpha)}{\beta},
\end{equation}
and the total flow rate at $P_{\mathrm{in}}^*$ is
\begin{equation}\label{Qstar}
Q_1+Q_2=Q_4+Q_5 = \frac{F_2(\alpha)}{\beta},
\end{equation}
where $F_1(\alpha)$ and $F_2(\alpha)$ are polynomial functions
of $\alpha$ with coefficients that depend on  the channel segment lengths, and $\beta$ is a property of the channel segment containing obstacles  (see the coefficient of the quadratic term in equation~\eqref{model5}). 
This dependence on $\beta$ highlights the importance of the Forchheimer effect for flow switching in the linking channel.
 
To analyze the flows through the system near the flow switching point, we consider a small deviation from the critical pressure by setting $P_{\mathrm{in}}=P_{\mathrm{in}}^* + \delta P_{\mathrm{in}}$. 
We then linearize the system by writing $\vec X = \vec X^* + \delta \vec X$, where $\vec X^*$ is the solution of
$\vec{G}(\vec X^*,P_{\mathrm{in}}^*) = 0$ and 
$\delta \vec X$ is a small deviation. This leads to
\begin{equation}
\mathrm{D} \vec G^* 
\cdot \delta \vec X + \frac{\partial \vec G}{\partial P_{\mathrm{in}}}\bigg\rvert_{(\vec X^*,P^*_{\mathrm{in}})} 
\delta P_{\mathrm{in}} =\vec 0,
\label{eq:16}
\end{equation}
where the quadratic terms in $\delta \vec X$ and $\delta P_{\mathrm{in}}$ have been removed and $\mathrm{D} \vec G^*$ is the Jacobian matrix of $\vec G$ evaluated at  $(\vec X^*,P_{\mathrm{in}}^*)$. 
The derivative of $\vec G$ with respect to pressure $P_{\mathrm{in}}$ is simply given by $\frac{\partial \vec G}{\partial P_{\mathrm{in}}} = \vec e_1+\vec e_2$, where $\vec e_j$ is the $j$-th coordinate unit vector. 
To verify that the flow through the linking channel indeed switches directions at $(\vec X^*,P_{\mathrm{in}}^*) $, we solve equation~\eqref{eq:16} for the variation in $Q_{3}$:
\begin{equation}
\delta Q_3 = - \vec e_5^T  {\mathrm{D} \vec G^*}^{-1} (\vec e_1+\vec e_2) \, \delta P_{\mathrm{in}},
\label{deltaQlink1}
\end{equation}
where $\mathrm{D} \vec G^*$ is verified to be invertible for the parameters we simulate.
Explicit calculation of equation~(\ref{deltaQlink1}) yields
\begin{equation}
\delta Q_{3} = - \frac{A(\alpha)\kappa}{B(\alpha) \kappa + C(\alpha)} \, \delta P_{\mathrm{in}},
\label{deltaQlink2}
\end{equation}
where $A(\alpha)$, $B(\alpha)$, and $C(\alpha)$ are polynomial functions of $\alpha$ with coefficients that only depend on the lengths of the channel segments. It can be shown that $A$, $B$, and $C$ are strictly positive if equation~(\ref{L1A}) is satisfied. We therefore observe that increasing the inlet pressure above the critical point $P^*_{\mathrm{in}}$ forces the fluid to flow in the negative direction ($\delta Q_{3} < 0$), whereas for $P_{\mathrm{in}} < P^*_{\mathrm{in}}$, the flow rate through the linking channel is positive, which indicates a switch in the direction of flow through the linking channel at $P^*_{\mathrm{in}}$.

\subsection{Flow switching under controlled total pressure}\label{S1.2}
 
\noindent
When the total pressure is controlled at the inlets, we can consider the inlets of the system in Supplementary Fig.~\ref{schem} to be directly connected to a common pressurized reservoir. We now take $P_{\mathrm{in}}$ to be the pressure of the reservoir and thus the total pressure at the inlets.
Then, the non-dimensional static pressure at the inlet of channel segment 1 can be expressed as $P_{\mathrm{in}} -  \frac{1}{2} {Q_{1}^2}$  and the static pressure at the inlet of channel segment 2 as $P_{\mathrm{in}} -  \frac{1}{2} {Q_{2}^2}$.
Now, the model near the flow switching point can be written as in equations~(\ref{model1})-(\ref{model8}), but with equations~(\ref{model1})-(\ref{model2}) replaced by
\begin{eqnarray}
 P_{\mathrm{in}} -  \frac{1}{2} {Q_{1}^2} - P_1 -  L_{1} Q_{1} &=& 0, \label{tpModel1}\\
 P_{\mathrm{in}} -  \frac{1}{2} Q_{2}^2 - P_2 -  L_{2} Q_{2} &=& 0. \label{tpModel2}
 \end{eqnarray}

We perform analysis similar to that done in section~\ref{S1.1}
and recover the same condition established in equation~(\ref{L1A}) for the existence of a (physical) solution for $P_{\mathrm{in}}^*>0$, which is now the total pressure at which a flow switch occurs. We also find the variation in $Q_3$  around $P_{\mathrm{in}}^*$ to be
\begin{equation}
\delta Q_{\mathrm{3}} = - \frac{\kappa}{\widetilde{A} \, \kappa + \widetilde{B}} \, \delta P_{\mathrm{in}},
\end{equation}
where $\widetilde{A}$ and $\widetilde{B}$ are functions that depend on $\alpha$, $\beta$, and the channel segment lengths,
but are both positive for the range of parameters we use. Therefore, we see that the flow through the linking channel changes direction at the critical point $(\vec X^*,P_{\mathrm{in}}^*)$, which is analogous to the result from equation~(\ref{deltaQlink2}) for the static pressure controlled case.

\section{Accounting for minor losses}\label{S2}

\noindent
The Reynolds number of the flows 
through the microfluidic system in Supplementary Fig.~\ref{schem} can be of the order of $1$ to $1000$ under the conditions of our study. Fluid inertia effects are therefore expected to be present, and additional pressure losses at channel junctions, also called \textit{minor} losses, should be considered.
The flow rates through channel segments 1 and 5 
are an order of magnitude higher than flow rates through the other channels and an extra  loss term should be added to equation~(\ref{model3}) to account for minor losses in segment 5 due to the junction with the linking channel. This minor loss is expected to scale linearly with the average flow velocity ($Q_5/w$ in dimensional values) 
at low $Re$ and quadratically 
at high $Re$.
Here, we use a general formulation to model the minor losses in segment 5, where the coefficient of the scaling factor (usually found empirically) depends on the ratio of the combining or diverging flows \cite{Crane1978}.
The pressure loss equation for channel segment 5, now including a minor loss term, is
\begin{equation} \label{minor}
\frac{P_1 - P_{\mathrm{out}}}{L_{5}} =  Q_{5} - K\bigg(\frac{Q_{3}}{Q_{1}}\bigg) \, f(Q_{5}),
\end{equation} 
where,
 as we note in the main text, 
 the scaling factor $f(Q_5)$ and the coefficient $K(Q_3/Q_1)$ are increasing functions for $P_{\mathrm{in}} \ge 0$  such that  $f(0)=K(0)=0$. 
The latter is consistent with the physical condition of having no minor losses when there is no flow through the linking channel.
The inclusion of this minor loss term in equation~(\ref{model3}) does not alter the condition in equation~\eqref{L1A} for $P_{\mathrm{in}}^*>0$ and the associated flow switch.
But minor losses are determinant for the emergence of Braess's paradox, as shown next, both when the static pressure and when the total pressure is controlled at the inlets.

\subsection{Condition for Braess's paradox under controlled static pressure}\label{S2.1}

\noindent
We modify our model for the static pressure controlled case by replacing equation~(\ref{model3}) with equation~(\ref{minor}). Now, the function $\vec G$ 
used to define our model in equation~\eqref{G} takes the form
\begin{equation}
\vec G(\vec X, P_{\mathrm{in}}) = \mathbf{A} \vec X + \vec B -   \frac{1}{12}  L_{4} \beta X_6^2 \, \vec e_4 + L_{5}
K\bigg(\frac{X_5}{X_3}\bigg) f(X_7) \, \vec e_3,
\end{equation}
where $\mathbf{A}$ is the matrix containing the coefficients of the linear terms in
 equations~(\ref{model1})-(\ref{model2}), equation~(\ref{minor}), and equations~(\ref{model5})-(\ref{model8}),
and $\vec{B}~=~(P_{\mathrm{in}},P_{\mathrm{in}},-P_{\mathrm{out}},-P_{\mathrm{out}},0,0,0)^T$ is a vector containing the imposed static pressures at the inlets and outlets.

The Jacobian of $\vec G$ at any point $\vec X$, with quadratic terms removed, reads
\begin{equation}
\mathrm{D} \vec G = \mathbf{A} -\frac{1}{6} L_{4} \beta X_6 \, \vec e_4 \otimes \vec e_6
+ L_{5} K\bigg(\frac{X_5}{X_3}\bigg) f'(X_7) \, \vec e_3 \otimes \vec e_7 
- L_{5} \frac{X_5}{X_3^2} K'\bigg(\frac{X_5}{X_3}\bigg) f(X_7) \, \vec e_3 \otimes \vec e_3 
+   \frac{L_{5}}{X_3} K'\bigg(\frac{X_5}{X_3}\bigg) f(X_7) \, \vec e_3 \otimes \vec e_5,
\end{equation}
where $\vec v\otimes \vec u\equiv\vec v\vec u^T$ indicates the outer product of $\vec v$  and  $\vec u$, and primes denote derivatives.
At the critical point $(\vec X^*, P^*_{\mathrm{in}})$, by construction, $Q_3 =X_5= 0$ and  $K(0)=0$, and thus the Jacobian reads
\begin{equation}
\label{linear}
\mathrm{D} \vec G^* = \mathbf{A} -\frac{1}{6} L_{4} \beta X_6^* \, \vec e_4 \otimes \vec e_6
+   \frac{L_{5}}{X_3^*} K'(0) f(X_7^*) \, \vec e_3 \otimes \vec e_5. 
\end{equation}
We have checked that this matrix is non-singular for the range of parameters used here. 

Following the notation in the main text,  we use $\Delta \mathcal{Q}=Q_{C} - Q_{D} $ to denote the difference between the total flow rates  ($Q_4+Q_5$) for the connected ($Q_{C}$) and disconnected ($Q_{D}$) system configurations. 
Similarly, we designate the difference in individual channel flow rates between the two configurations by $\Delta Q_i=Q_{i,C} - Q_{i,D}$. 
Under a small variation $\delta P_{\mathrm{in}}$ around the flow switching point $P_{\mathrm{in}}^*$, at which $Q_{C} =    Q_{D}$ and $Q_{i,C} =  Q_{i,D} $, we have
 \begin{equation}
 \Delta \mathcal{Q} = \delta Q_{C}  -  \delta  Q_{D},
 \label{Qtot}
  \end{equation}
  \begin{equation}
 \Delta Q_i =  \delta Q_{i,C}  -  \delta  Q_{i,D}.
 \label{Qi}
 \end{equation}  
To find $\Delta \mathcal{Q}$, we use the fact that $\delta Q_C = \delta Q_{4,C} + \delta Q_{5,C}$ and that $\delta Q_D$  
can be calculated by taking 
the limit of $\delta Q_{C}$ when $\kappa \to 0$ (i.e., the limit of infinite resistance for the linking channel). After explicit calculation using equation~(\ref{linear}), we find
 \begin{equation}
 \Delta \mathcal{Q} = \kappa  \delta P_{\mathrm{in}} \frac{b_1(\alpha) - b_2(\alpha)   \, K'(0) \beta f(\frac{a_1(\alpha)}{\beta}) }{a_2(\alpha)+a_3(\alpha) \kappa + a_4(\alpha) \kappa K'(0) \beta f(\frac{a_1(\alpha)}{\beta})},
 \label{deltaQtot}
  \end{equation}
 where the $a_i(\alpha)$ and $b_i(\alpha)$ are polynomials of $\alpha$ with coefficients that only depend on the channel segment lengths and are positive when equation~(\ref{L1A}) is satisfied. We therefore conclude that, when equation~(\ref{L1A}) is satisfied, Braess's paradox occurs  for $\delta P_{\mathrm{in}}>0$, as observed in our Navier-Stokes simulations and experiments,  if and only~if
 \begin{equation}
 K'(0) \beta f\bigg(\frac{a_1}{\beta}\bigg) > \frac{b_1}{b_2},
 \label{paradoxCond}
 \end{equation}
 where the presence of $\beta$ and $K'(0)$ highlight the crucial role of nonlinearity and minor losses.
 Equation \eqref{paradoxCond} indicates that if $Q_5$ is sensitive enough to the flow through the linking channel (i.e., $K'(0)$ is large enough), then the paradox will manifest itself.
 The compound interaction between the nonlinearity arising from the obstacles and minor losses is further illustrated by the relative magnitude of $\Delta \mathcal{Q}/Q_C$ near $P_{\mathrm{in}}^*$, where $Q_C = F_2(\alpha)/\beta$ from equation~(\ref{Qstar}).
  
The difference in the total flow rate, $\Delta \mathcal{Q}$, can also be broken down into the differences in $Q_4$ and $Q_5$. We find 
\begin{equation}
 \Delta Q_4 = -\kappa  \delta P_{\mathrm{in}} \frac{c_1(\alpha) }{c_2(\alpha)+c_3(\alpha) \kappa + c_4(\alpha) \kappa K'(0) \beta f(\frac{a_1(\alpha)}{\beta})},
 \label{DeltaQ4}
  \end{equation}
  \begin{equation}
  \Delta Q_5 = \kappa  \delta P_{\mathrm{in}} \frac{g_1(\alpha) -g_2(\alpha)K'(0) \beta f(\frac{a_1(\alpha)}{\beta})}{g_3(\alpha)+g_4(\alpha) \kappa + g_5(\alpha) \kappa K'(0) \beta f(\frac{a_1(\alpha)}{\beta})},
 \label{DeltaQ5}
  \end{equation}
where, similarly, the $c_i(\alpha)$ and $g_i(\alpha)$ are polynomials of $\alpha$ with coefficients that only depend on the channel segment lengths and are positive when equation~(\ref{L1A}) is satisfied. Equations~(\ref{DeltaQ4})-(\ref{DeltaQ5}) show how minor losses impact the flow rates at each of system outlets (as discussed in section~\ref{S2.3}).

 \subsection{Condition for Braess's paradox under controlled total pressure}\label{S2.2}
\noindent
For the scenario in which total pressure is controlled at the inlets, we again substitute equation~(\ref{minor}) for equation~(\ref{model3}) to define our model in equation~(\ref{G}) with $\vec{G}$ now in the form 
\begin{equation}
\vec G(\vec X, P_{\mathrm{in}}) = \mathbf{\widetilde{A}} \vec X + \vec{\widetilde{B}} -   \frac{1}{12}  L_{4} \beta X_6^2 \, \vec e_4 - \frac{1}{2}X_3^2 \, \vec e_1  - \frac{1}{2}X_4^2 \, \vec e_2+ L_{5}
K\bigg(\frac{X_5}{X_3}\bigg) f(X_7) \, \vec e_3,
\end{equation}
where matrix $\mathbf{\widetilde{A}} = \mathbf{A}$ includes the coefficients of the linear terms in equations~(\ref{tpModel1})-(\ref{tpModel2}), equation~(\ref{minor}), and equations~(\ref{model5})-(\ref{model8}), and vector $\vec{\widetilde{B}}~=~(P_{\mathrm{in}},P_{\mathrm{in}},-P_{\mathrm{out}},-P_{\mathrm{out}},0,0,0)^T$ accounts for the total pressure at the inlets and static pressure at the outlets.
 
At the critical point $(\vec X^*, P^*_{\mathrm{in}})$, again $K(0)=0$, and the Jacobian reads
\begin{equation}
\mathrm{D} \vec G^* = \mathbf{\widetilde{A}} -\frac{1}{6} L_{4} \beta X_6^* \, \vec e_4 \otimes \vec e_6 - X_3^* \, \vec e_1 \otimes \vec e_3 - X_4^* \, \vec e_2 \otimes \vec e_4
+   \frac{L_{5}}{X_3^*} K'(0) f(X_7^*) \, \vec e_3 \otimes \vec e_5. 
\end{equation}
Performing similar analysis to that done in section~\ref{S2.1}, we find 
 \begin{equation}
 \Delta \mathcal{Q} = \kappa  \delta P_{\mathrm{in}} \frac{\widetilde{b}_1 - \widetilde{b}_2   \, K'(0) f(\widetilde{a}_1) }{\widetilde{a}_2+\widetilde{a}_3 \kappa + \widetilde{a}_4 \kappa K'(0) f(\widetilde{a}_1)},
 \label{deltaQtot_tp}
  \end{equation}
 where  $\widetilde{a}_i$ and $\widetilde{b}_i$ are functions that can depend on channel lengths $L_i$, parameter $\alpha$, and parameter $\beta$, and they are positive for the range of parameters we consider. We therefore conclude that Braess's paradox occurs for $\delta P_{\mathrm{in}}>0$ if
 \begin{equation}
 K'(0) f(\widetilde{a}_1) > \frac{\widetilde{b}_1}{\widetilde{b}_2},
 \label{paradoxCond_tp}
 \end{equation}
 similarly to the case when static pressure is controlled.

\subsection{Prediction of Braess's paradox in model without minor losses}\label{S2.3}
We now elaborate on the need to account for minor losses in order to predict Braess's paradox as observed in our experiments and simulations.
The model prediction of Braess's paradox near $P_{\mathrm{in}}^*$, while neglecting minor losses and under static pressure control, can be found directly from equation (\ref{deltaQtot}) by removing the minor loss terms. Specifically, we have 
\begin{equation}
\Delta \mathcal{Q} = \kappa  \delta P_{\mathrm{in}} \frac{b_1(\alpha)}{a_2(\alpha)+a_3(\alpha) \kappa}.
\label{deltaQtot_noML}
\end{equation}
Therefore, the paradox is predicted to exist for $\delta P_{\mathrm{in}}<0$. 
We show in Supplementary Fig.~\ref{paradox_noML}a the predictions of $Q_3$ and $\Delta \mathcal{Q}$ for the network used in Fig.~\ref{fig3} when minor losses are not included in the model by numerically solving equations~(\ref{model1})-(\ref{model8}). Braess's paradox is predicted to occur at $P_{\mathrm{in}}$ below $P^*_{\mathrm{in}}$ with a relative magnitude of less than $0.1\%$. We also show in Supplementary Fig.~\ref{paradox_noML}b the predicted difference in each $Q_4$ and $Q_5$ between the connected and disconnected systems. Above $P_{\mathrm{in}}^*$, removing the linking channel results in a small increase in $Q_4$ and a slightly larger decrease in $Q_5$, thus resulting in a small net decrease in the total flow rate $(Q_4+Q_5)$. This directly contrasts with the simulation results in Fig.~\ref{fig3}, where the paradox is observed for $P_{\mathrm{in}}$ above $P_{\mathrm{in}}^*$ in which the removal of the linking channel results in a significant \textit{increase} in $Q_5$. 

To further determine how the inclusion of minor losses in the model alters the prediction of the paradox occurring above or below $P_{\mathrm{in}}^*$, we consider the difference in $Q_4$ and $Q_5$, individually, between the connected and disconnected system configurations near $P_{\mathrm{in}}^*$. By removing minor loss terms from equations~(\ref{DeltaQ4})-(\ref{DeltaQ5}), we find 
\begin{equation}
 \Delta Q_4 = -\kappa  \delta P_{\mathrm{in}} \frac{c_1(\alpha) }{c_2(\alpha)+c_3(\alpha) \kappa},
 \label{DeltaQ4_noML}
  \end{equation}
  \begin{equation}
  \Delta Q_5 = \kappa  \delta P_{\mathrm{in}} \frac{g_1(\alpha)}{g_3(\alpha)+g_4(\alpha) \kappa}.
 \label{DeltaQ5_noML}
  \end{equation}
Several conclusions follow immediately from equations~(\ref{DeltaQ4_noML})-(\ref{DeltaQ5_noML}). First, for a small increase in the driving pressure above $P_{\mathrm{in}}^*$ (i.e., $\delta P_{\mathrm{in}}>0$), we predict $\Delta Q_4$ to be negative whether minor losses are accounted for or not. This implies that removing the linking channel at $P_{\mathrm{in}}$  slightly above $P_{\mathrm{in}}^*$, leads to an increase in $Q_4$. 
Second, with minor loss terms neglected, we expect $\Delta Q_5$ to be positive for  $\delta P_{\mathrm{in}}>0$. This is in accordance with Supplementary Fig.~\ref{paradox_noML}b, in which removing the linking channel at $P_{\mathrm{in}}>P_{\mathrm{in}}^*$ results in a decrease in $Q_5$. However, this contrasts with the result in equation~(\ref{DeltaQ5}), where we see that $\Delta Q_5$ is negative for $\delta P_{\mathrm{in}}>0$ if $K'(0) \beta f(\frac{a_1(\alpha)}{\beta})>g_1(\alpha)/g_2(\alpha)$. That is, if minor losses are large enough, removing the linking channel (and thus removing the minor losses themselves) can result in an increase in both $Q_4$ and $Q_5$ for $P_{\mathrm{in}}>P_{\mathrm{in}}^*$, which is consistent with our simulation and experimental results.

\begin{figure}[htb] 
\centering 
\includegraphics[width=0.95\textwidth]{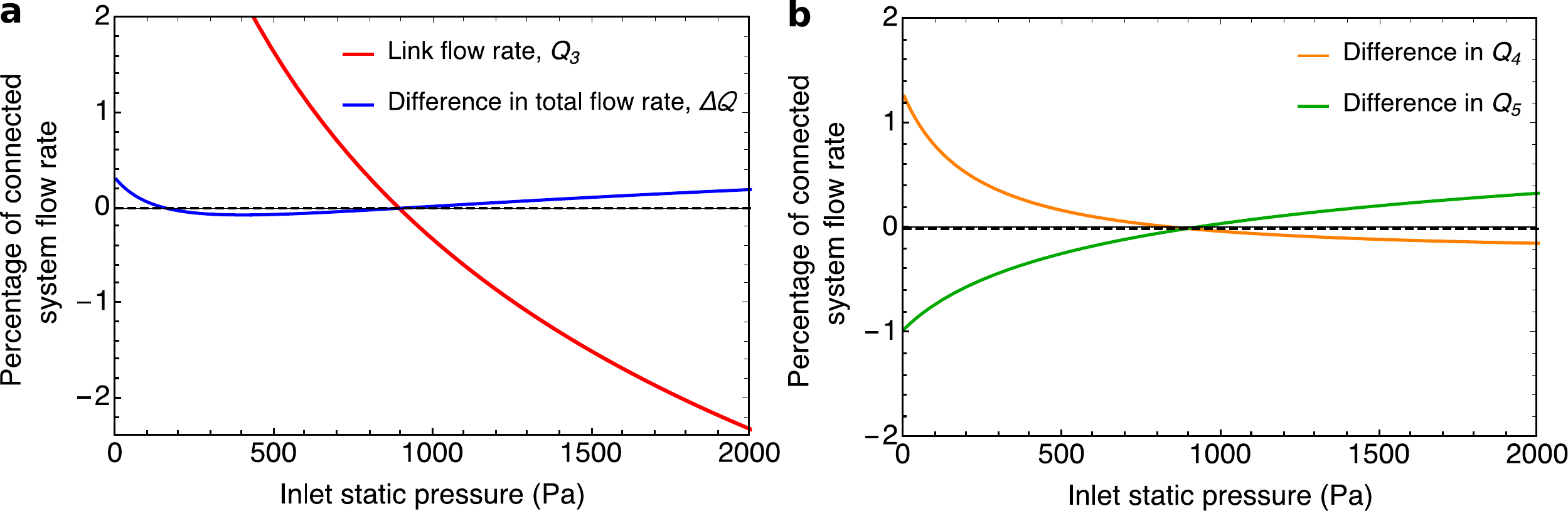}
\caption{\textbf{Model prediction of Braess's paradox without minor losses.}
\textbf{a}, Flow rate through the linking channel $Q_3$ and difference in total flow rate $\Delta \mathcal{Q} = Q_C - Q_D$ between the connected and disconnected system configurations as a percentage of $Q_C$. Braess's paradox is only predicted for a range of $P_{\mathrm{in}}$ below $P_{\mathrm{in}}^*$ (specifically, the range where $\Delta \mathcal{Q}$ is negative). 
\textbf{b}, Differences in $Q_4$ and $Q_5$ between the connected and disconnected configurations. Positive values indicate that the quantity is larger for the connected system configuration. Above $P_{\mathrm{in}}^*$, removing the linking channel causes $Q_4$ to increase by a small amount and $Q_5$ to decrease by a slightly larger amount. Hence, Braess's paradox is not predicted for $P_{\mathrm{in}}>P_{\mathrm{in}}^*$.
The dimensions of the channels used here are the same as those used in Fig.~\ref{fig3}. For the linking channel, we estimate the hydraulic resistance as $12\mu L_3/w^3$ and take $\gamma=1.03$.
}
\label{paradox_noML}
\end{figure}

\newpage

\section{Supplemental simulation results for channels with obstacles, flow switching, and Braess's paradox}\label{S3}

In the following sections, we provide results from fluid dynamics simulations on the pressure-flow relation for channels with obstacles, the verification of model predictions for flow switching, and the manifestation of Braess's paradox under different boundary conditions.

\subsection{Flows through channels with obstacles }\label{S3.1}

In Supplementary Fig.~\ref{fig5SI}a, we show how the relation between $\Delta P$ and $Re$   for a straight channel changes when obstacles are present. The nonlinearity of the relation increases with the number of obstacles and, in the main text, we relate this observed nonlinearity to the Forchheimer effect commonly found in porous media. One of the properties of the Forchheimer relation in equation~(\ref{eq2}) is that the coefficients $\alpha$ and $\beta$ only depend on the geometric structure of the system and not on properties of the working fluid. We verify that this property also carries over to our system in Supplementary Fig.~\ref{fig5SI}b. First, we fit the relation between $Re$ and $-\Delta P / Re$ to determine $\alpha$ and $\beta$ for a channel with ten obstacles. Then, we use these coefficients to predict the same relation for the same channel when a working fluid with a different viscosity is used. The excellent agreement between the prediction and the simulations confirms that $\alpha$ and $\beta$ are not dependent on the fluid properties. 
We also observe that as the flow rate is reduced to an $Re$ below $O(1)$, inertial effects become negligible and the relation between $Re$ and $-\Delta P / Re$ plateaus to a constant value (Supplementary Fig.~\ref{fig5SI}c).

\begin{figure}[hbt] 
\centering 
\includegraphics[width=0.99\textwidth]{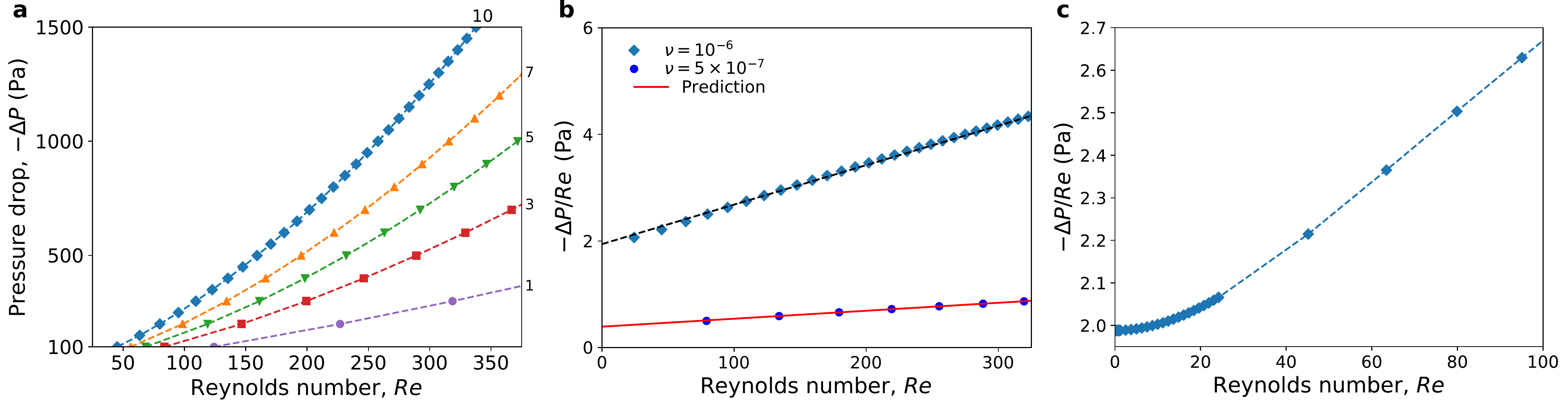}
\caption{\textbf{Nonlinearty in flow through a channel with obstacles.} 
\textbf{a}, Simulation results of the relation between Reynolds number and pressure loss  
 for a channel with varying numbers of cylindrical obstacles (indicated next to each curve).
The simulations are performed with a water-like fluid (kinematic viscosity $\nu =10^{-6}\,$m$^2$/s).
\textbf{b}, Relation between $Re$ and $-\Delta P/Re$ for the channel with ten obstacles (same data shown in \textbf{a}) is fit with a straight line (dashed line). Simulation results are indicated by symbols. The fitted parameters, $\alpha$ and $\beta$, are used to predict the same relation for flow through the same channel but for a fluid with $\nu =5\times 10^{-7}\,$m$^2$/s (continuous line).
\textbf{c}, Relation between $Re$ and $-\Delta P/Re$ for the ten-obstacle channel in \textbf{a} at lower values of $Re$.
The flattening of the relation as $Re$ approaches zero shows that the pressure-flow relation is approximately linear for $Re\le O(1)$.
In all panels, the lengths of the channels are $1.25\,$cm,
and the
fitted parameters in \textbf{b} are $\alpha=1.62\times 10^8\,$m$^{-2}$ and $\beta =570\,$m$^{-1}$.
}
\label{fig5SI}
\end{figure}

\newpage

\subsection{Verification of model flow switching predictions}\label{S3.2}
\begin{figure*}[htb] 
\centering 
\includegraphics[width=0.99\textwidth]{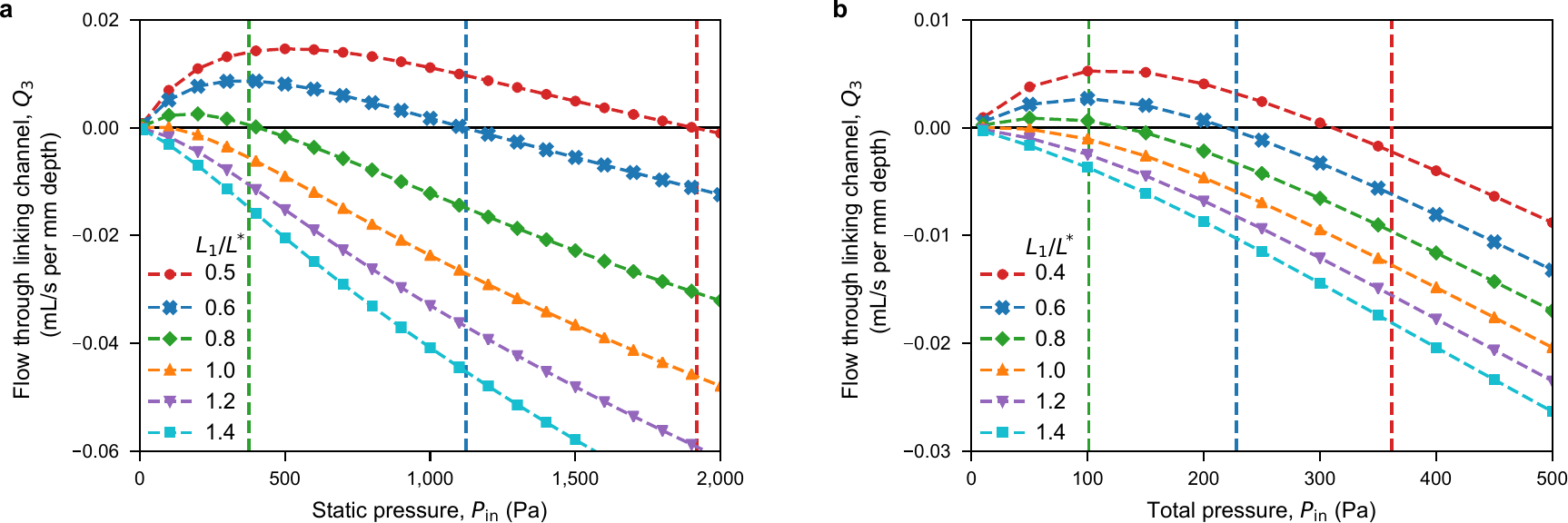}
\caption{\textbf{Model prediction of flow switching.} \textbf{a}, \textbf{b}, Simulated flow rate through the linking channel (symbols) for different values of $L_{1}$ when controlling static pressure (\textbf{a}) and total pressure (\textbf{b}). In agreement with the prediction in equation~(\ref{eq3}), 
a flow switch occurs only when  $L_{1}/L^* < 1$.  
The vertical lines indicate the model prediction of $P_{\mathrm{in}}^*$ for the curves with corresponding color. The values of the free parameter $\gamma$ used for the static and total pressure controlled cases are 1.02 and 0.99, respectively, and the dimensions of the channel segments are:  
$L_2 =3.0\,$cm, $L_3 =0.1\,$cm, $L_4 =1.25\,$cm,  $L_5 =1.4\,$cm, and $L^*=0.995\,$cm (calculated using the value for $\alpha$ found in Supplementary Fig.~\ref{fig5SI}b).
}
\label{figS4}
\end{figure*}

Supplementary Fig.~\ref{figS4} shows
validation of the model prediction in equation~\eqref{eq3} for flow switching through comparison with results from simulations of the Navier-Stokes equations.
The pressure at which the flow switch occurs tends to $0$ as $L_{1}/L^*$ approaches $1$ from below, and  $Q_{3} $ is negative for any positive inlet pressure when $L_{1}/L^* > 1$. This result is found both when the static (Supplementary Fig.~\ref{figS4}a) and when the total (Supplementary Fig.~\ref{figS4}b) pressure is controlled and shows that the analytic model agrees quantitatively with our findings above.
In predicting the precise pressure at which the flow switch occurs, the conductivity of the linking channel, $\kappa$, is inconsequential. However, the prediction is  sensitive to the parameter $\gamma$. Thus, we treat $\gamma$ as a fitting parameter to predict the pressure at which the flow switch occurs, as indicated by vertical lines in the figure. The resulting predictions are in excellent agreement with simulations when static pressure is controlled (Supplementary Fig.~\ref{figS4}a). The predictions when total pressure is controlled are less accurate (Supplementary Fig.~\ref{figS4}b), likely as a result of not including entrance length effects in the model, which would account for the development 
of the  parabolic velocity profile characteristic of Poiseuille flow.

\noindent
Another important model prediction for the static pressure controlled case is that when all pressure loss equations are linear (i.e., $\beta = 0$), there is no strictly positive pressure $P_{\mathrm{in}}$ at which 
the flow rate through the linking channel is zero (i.e., $Q_{3} =0$, which would indicate a flow switching point). This implies that nonlinearity is necessary for the observed flow switching effect. We test this prediction using simulations of the connected system configuration but with the obstacles removed from channel segment 4 in Supplementary Fig.~\ref{schem}.  
Supplementary Fig.~\ref{0cyl}a,b shows the flow rate through the linking channel both when static pressure and when total pressure is controlled at the inlets. We see that, indeed, no switch in the direction of flow through the linking channel is observed. 

\begin{figure}[htb] 
\centering 
\includegraphics[width=0.95\textwidth]{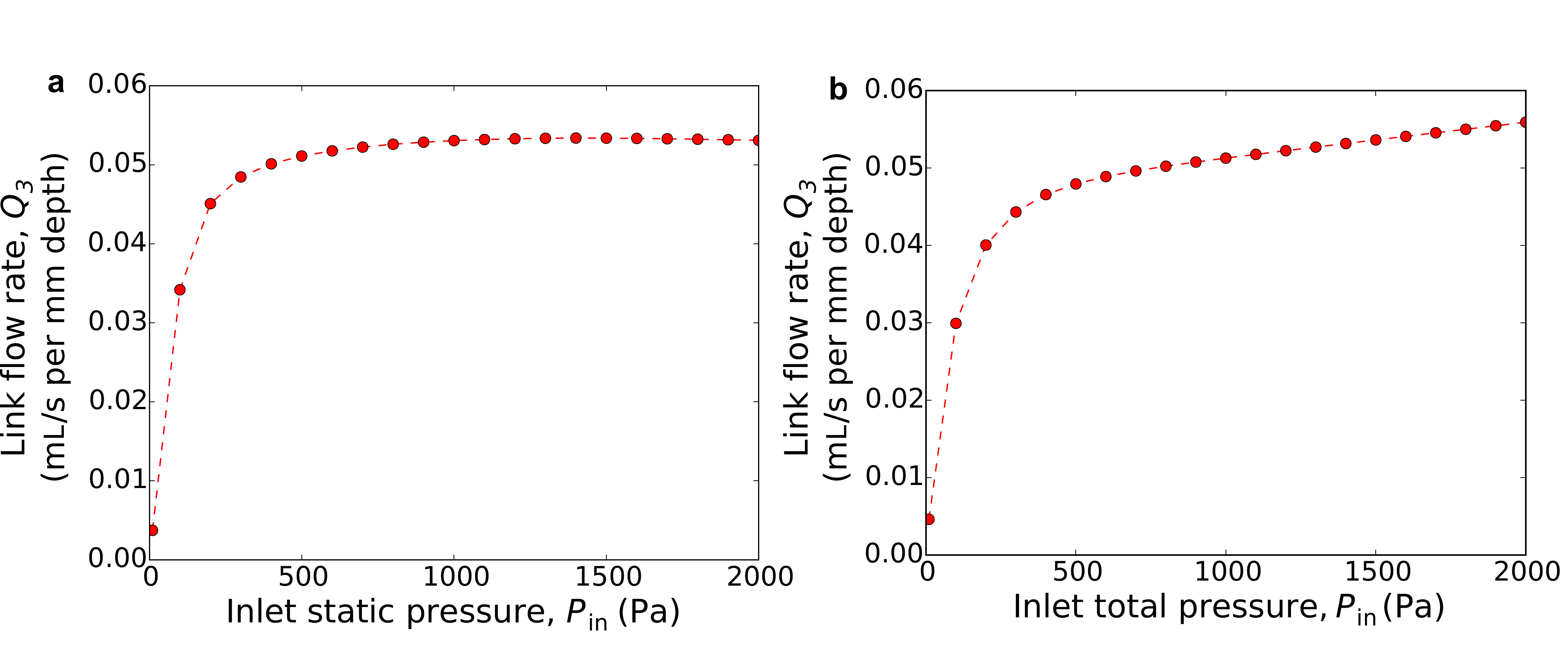}
\caption{\textbf{Flow through linking channel for system without obstacles.} \textbf{a}, \textbf{b}, Simulation results for flow rate through the linking channel in the absence of obstacles in channel segment 4 when the controlled variable $P_\mathrm{in}$ is the static pressure (\textbf{a}) and total pressure (\textbf{b}).
We observe no change in flow direction for either case for the range of pressures considered. Following the labels in Supplementary Fig.~\ref{schem}, the dimensions of the channels used in \textbf{a} are $L_1 =0.17$, $L_2 =0.85$, $L_3 =0.1$, $L_4 =1.25$, and $L_5 =1.0$; in \textbf{b} the dimensions used are $L_1 =0.25$, $L_2 =3.0$, $L_3 =0.1$, $L_4 =1.25$, and  $L_5 =1.4$ (all in units of cm).
}
\label{0cyl}
\end{figure}

\subsection{Braess's paradox under controlled static pressure}\label{S3.3}

In analyzing  the extent of Braess's paradox as a function of $P_{\mathrm{in}}$, two representations of the data are natural. 
One is the representation adopted in Fig.~\ref{fig3} of the main text, in which the flow rates are shown as percentages of
the total flow rate for the connected system configuration, $Q_C$. This accommodates the fact that the total flow rate varies significantly 
with pressure, and therefore expresses the relative magnitude of the paradox. 
In Supplementary Fig.~\ref{paradoxSP}, we visualize the same data in dimensional values of the flow rates. This representation is useful,
 for example, for confirming that $Q_3$ initially increases as a function of $P_{\mathrm{in}}$, which is not directly evident from Fig.~\ref{fig3}.

\begin{figure}[h!!] 
\centering 
\includegraphics[width=0.5\textwidth]{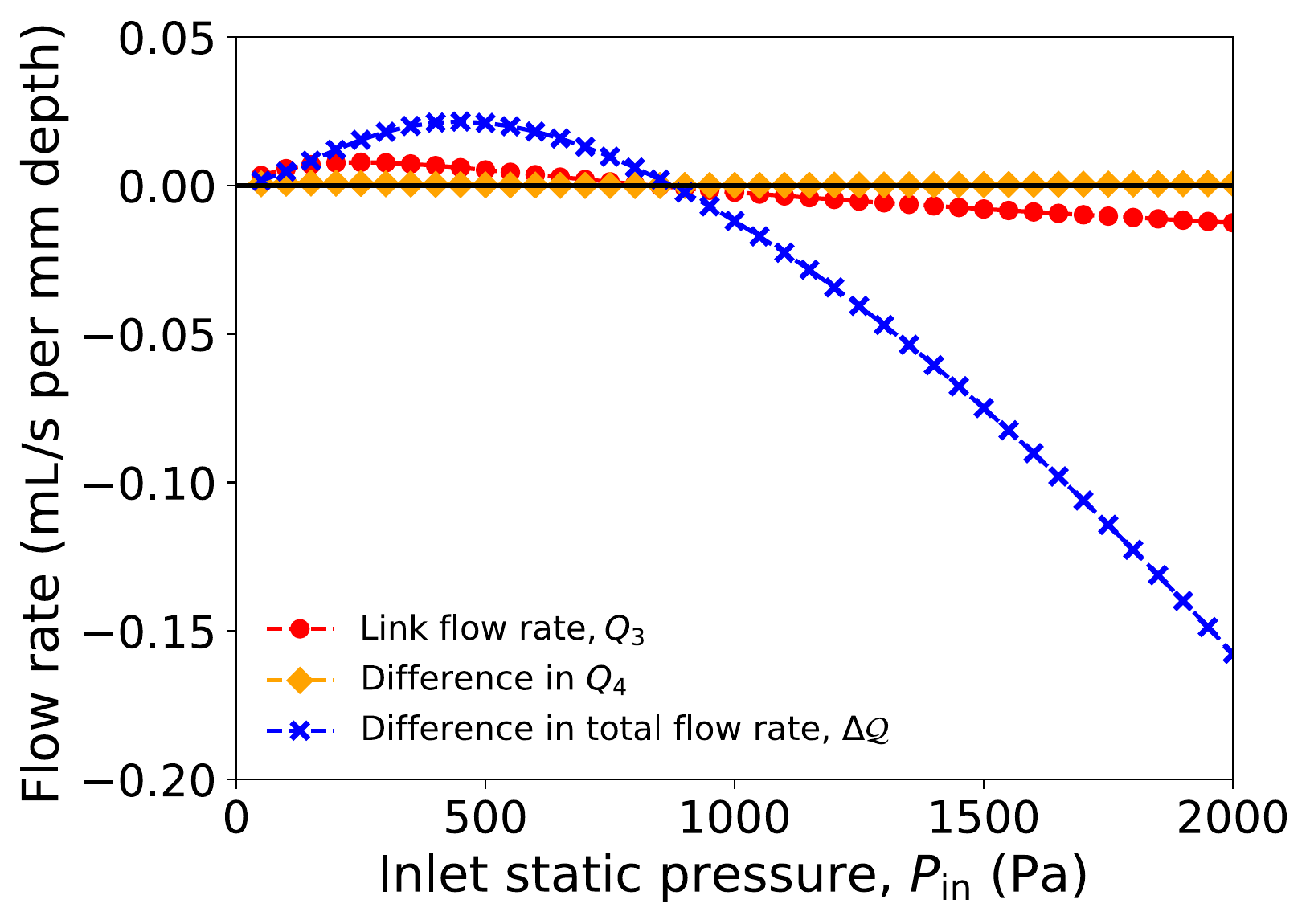}
\caption{\textbf{Braess's paradox under controlled static pressure.} Simulation results for the connected and disconnected system configurations in dimensional form. This figure corresponds to Fig.~\ref{fig3}, which shows the same data plotted as a percentage of the total flow rate $Q_C$.
}
\label{paradoxSP}
\end{figure}

We also consider how the geometry of the linking channel may influence the extent of Braess's paradox in the system presented in Fig.~\ref{fig1}a. In Supplementary Fig.~\ref{paradox_slanted}, we show the difference in the total flow rate through the connected and disconnected system configurations for networks in which the linking channel joins the parallel channels at different angles. This geometric change to the system slightly shifts the critical switching pressure $P_{\mathrm{in}}^*$ (which may be accounted for in the model by adjusting the value of $\gamma$ in equation~(\ref{model6})), but does not alter the emergence of the paradox near $P_{\mathrm{in}}^*$. Moreover, at higher pressures, the magnitude of the paradox can even be enhanced when the junctions with the linking channel deviate from a straight T-junction.

\begin{figure}[h!!] 
\centering 
\includegraphics[width=0.5\textwidth]{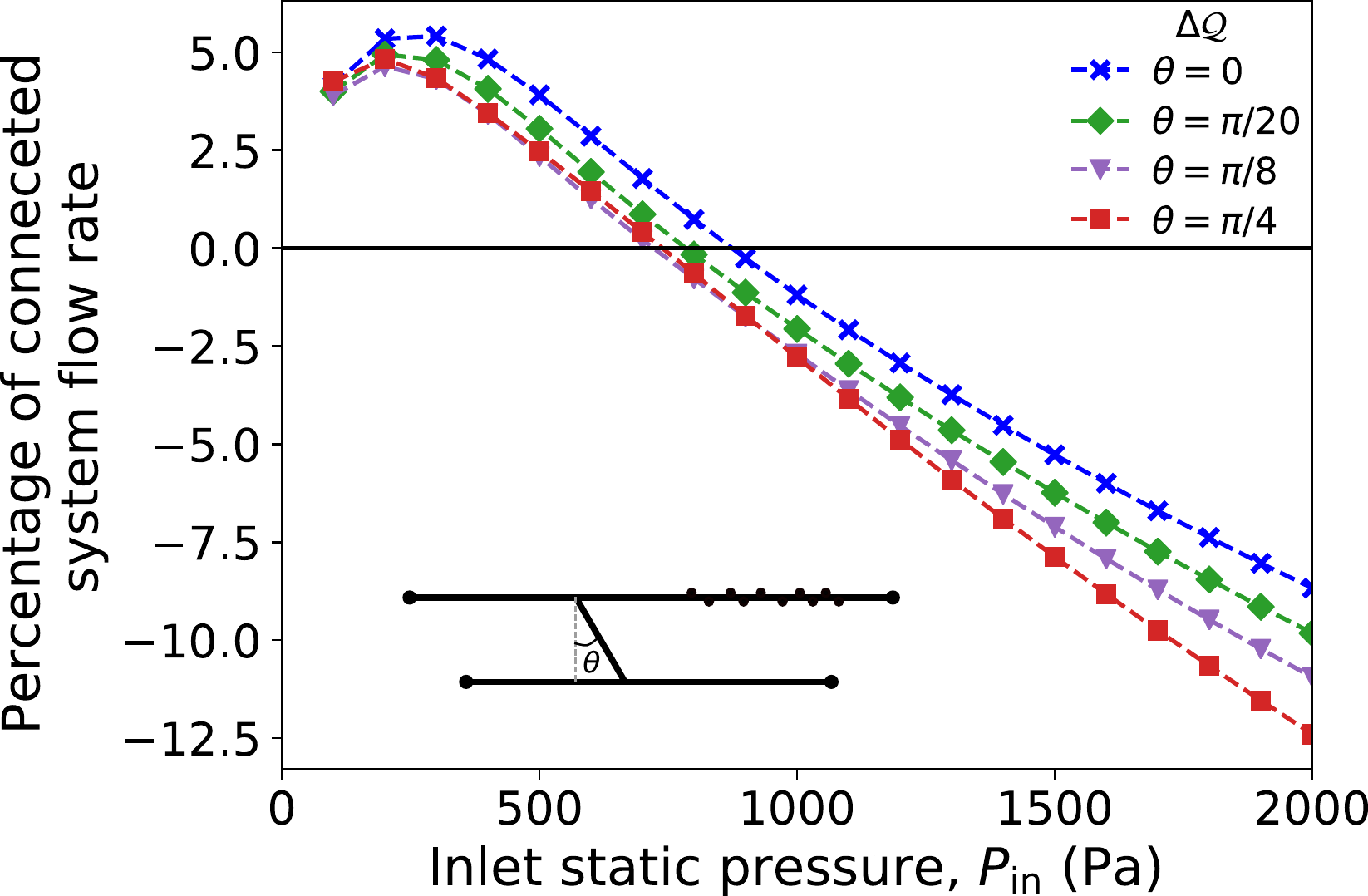}
\caption{\textbf{Braess's paradox for different linking channel geometries.} 
Comparison of flows between the connected and disconnected system configurations for controlled static pressure when the linking channel connects the two parallel channels at different angles. The inset schematic illustrates the network structure and defines the angle $\theta$. Crossing points on the $x$-axis were verified to correspond to the critical switching pressure for flow through the linking channel (not shown). 
The case $\theta=0$ is the same system as in Fig.~\ref{fig3}, and the lengths of the channel segments $L_1, L_2, L_4$, and $L_5$ are the same for each $\theta$. The length of the linking channel is $(0.1/\cos \theta)\,$cm. 
}
\label{paradox_slanted}
\end{figure}

\newpage

\subsection{Braess's paradox under controlled total pressure}\label{S3.4}
 
\begin{figure}[h!!] 
\centering 
\includegraphics[width=0.95\textwidth]{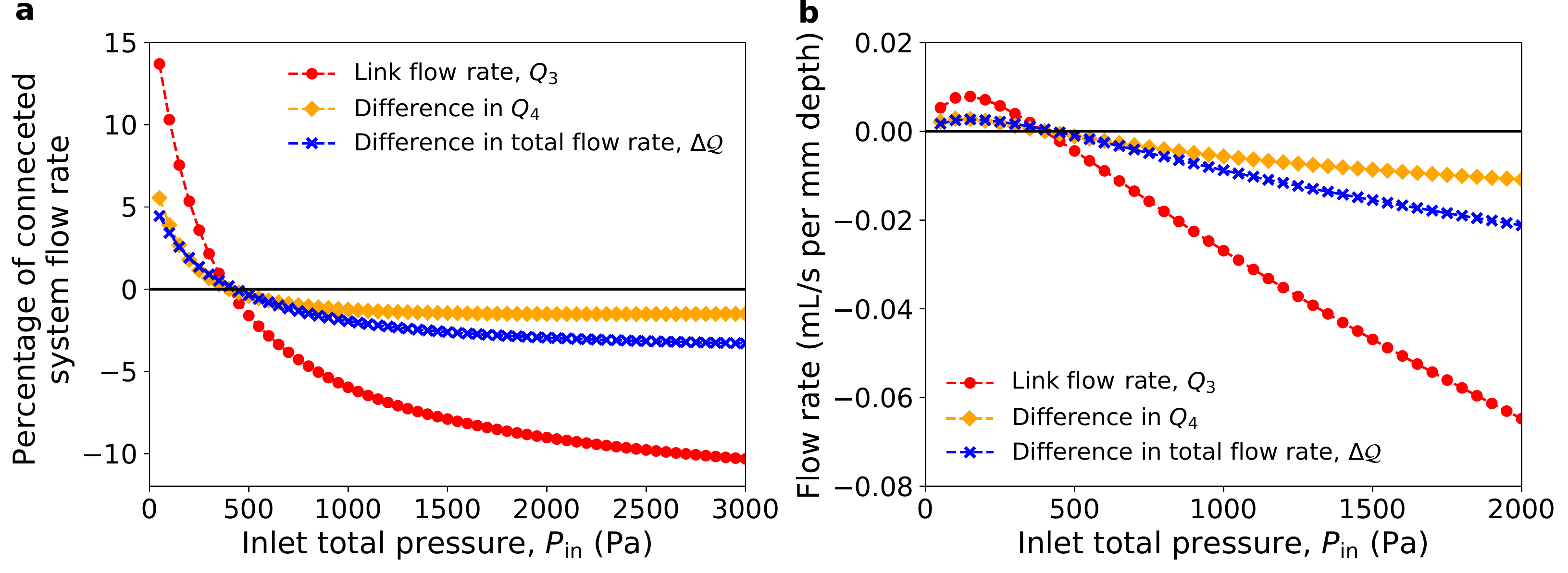}
\caption{ \textbf{Braess's paradox under controlled total pressure.} Comparison of flow rates through the connected and disconnected configurations when the total pressure is controlled at the inlets. \textbf{a}, Flow rates  plotted as a percentage of the total flow rate $Q_C$. \textbf{b}, Dimensional values of the flow rates shown in \textbf{a}. 
Panels \textbf{a} and \textbf{b} are the counterparts of Fig.~\ref{fig3} and Supplementary Fig.~\ref{paradoxSP}, respectively, where static pressure is controlled. The dimensions of the channels are the same as for Supplementary Fig.~\ref{figS4} with $L_1=0.25\,$cm.}
\label{paradoxTP1}
\end{figure}

\noindent
In the main text, the discussion on the simulation results focuses mainly on the scenario in which a common static pressure is controlled at the inlets. This is physically achievable by connecting a pressure regulator to each inlet and then connecting the system to a pressurized reservoir.  
In our experiments, the system channels directly connect to the pressurized reservoir without intermediate pressure regulators. In this case, the total pressure at the inlets is being controlled and is indeed equal to the pressure of the reservoir.  Our simulations show that the results carry over, as shown in Supplementary Fig.~\ref{figS4}, for the occurrence of flow switching both when static pressure is controlled and when total pressure is controlled.
We performed additional simulations to verify that  
the same holds true for Braess's paradox itself. Supplementary Fig.~\ref{paradoxTP1}
 confirms that the paradox persists in the total pressure controlled case and that the onset of the paradox occurs at the flow switching point $P_{\mathrm{in}}^*$, just as was seen in the static pressure controlled case. 
In addition, there do exist some specific differences between the static and total pressure controlled scenarios. 
When static pressure is controlled, there is little difference in the flow rate through the channel with obstacles, $Q_4$, between the cases in which the linking channel is open and closed. 
However, when total pressure is controlled, $Q_4$ and $Q_5$ increase in approximately equal magnitude
when the linking channel is closed for $P_{\mathrm{in}}>P_{\mathrm{in}}^*$, as was observed in our experiments (Fig.~\ref{fig4}c,d).
Also, as a percentage of $Q_C$, the flow rate through the linking channel is significantly larger and the magnitude of the paradox is smaller when total pressure is controlled than when static pressure is controlled (Supplementary Fig.~\ref{paradoxTP1}a). The latter point makes the conclusions in the main text stronger, since our experiments verified the predicted Braess paradox effect for the pressure boundary conditions under which the effect is weaker.

\newpage
\section{Prediction of negative and positive conductance transitions}\label{S4}

\begin{figure}[htb] 
\centering
\includegraphics[width=0.99\columnwidth]{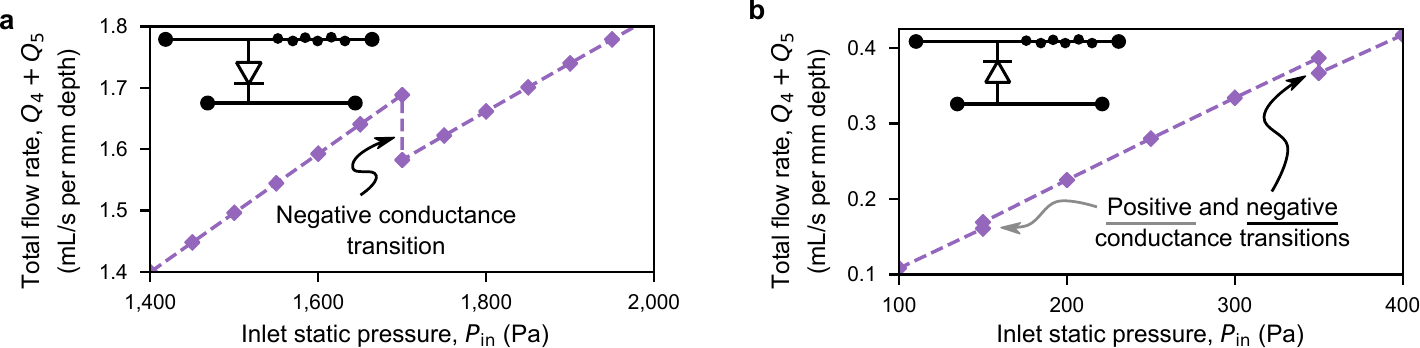}
\caption{\textbf{Negative and positive conductance transitions.} 
 \textbf{a}, \textbf{b}, Simulation results for the system with an offset fluidic diode incorporated into the linking channel, where the polarity of the diode is indicated in the inset network schematics. As the pressure $\Delta P_{21}$ passes a threshold value (positive in  \textbf{a} and negative in \textbf{b}), the system passively transitions from the connected to the disconnected configuration. For one polarity of the diode, this results in a negative conductance transition for  $P_{\mathrm{in}}>P_{\mathrm{in}}^*$ (\textbf{a}), whereas for the opposite polarity it results in a positive and a negative conductance transition for  $P_{\mathrm{in}}<P_{\mathrm{in}}^*$ (\textbf{b}). 
}
\label{diodeFig}
\end{figure}

Here, we consider how our observation of Braess's paradox and flow switching in the system presented in Fig.~\ref{fig3} may be harnessed by using an offset fluidic diode, which can be idealized as closed for pressure differences below a predefined threshold and  open above the threshold. 
Supplementary Fig.~\ref{diodeFig} shows results from simulations that incorporate one such diode with each of the two polarizations into the linking channel, where the state of the diode (open/closed) is governed by the pressure difference across the linking channel, $\Delta P_{21}$. 
In both cases, controlling the driving pressure can induce negative fluidic conductance transitions, where an \textit{increase} in $P_{\mathrm{in}}$ leads to an abrupt {\it decrease} in the total flow rate (and a decrease in $P_{\mathrm{in}}$ leads to an abrupt increase in flow rate). 
A negative conductance transition is predicted for any positive diode threshold (Supplementary Fig.~\ref{diodeFig}a) and another one is predicted for a negative diode threshold between zero and the observed minimum of $\Delta P_{21}$ (Supplementary Fig.~\ref{diodeFig}b). 
In the latter case, a small positive conductance transition is also predicted, which follows from the non-monotonic behavior of the flow rate $Q_3$ (and thus of $\Delta P_{21}$); the flow rate $Q_3$ initially increases and then decreases (i.e., $\Delta P_{21}$ passes through a minimum) as $P_{\mathrm{in}}$ is increased from $0$ to $P_{\mathrm{in}}^*$ (Supplementary Fig.~\ref{paradoxSP}). 
Note that the pressure difference $\Delta P_{21}$, and thus the opening and closing of the diode, is indirectly controlled  by varying  $P_{\mathrm{in}}$.
These transitions can be seen as a consequence of flow switching and Braess's paradox, in which the transition from the connected to the disconnected configuration is passive.

\newpage

\section{Supplemental experimental results for channels with obstacles, flow switching, and Braess's paradox}\label{S5}
The observed nonlinear pressure-flow relation for a channel containing obstacles is essential for programming a flow switch in the linking channel. To confirm inertial effects in the flow around the obstacles as the source of the nonlinearity, we experimentally measure the pressure-flow relation for channels constructed from materials with higher and lower rigidity than the PDMS composition used in the experiments presented in the main text. In Supplementary Fig.~\ref{ExpFlex3d}, we show the resulting relations between $Re$ and $-\Delta P/Re$ for channels fabricated from Flexdym and SU-8 photoresist (both with and without obstacles), as well as extended data from the PDMS channels presented in Fig.~\ref{fig2}e,f. The Flexdym and SU-8 photoresist channels both have approximately the same geometry and dimensions as the PDMS channels.
The SU-8 photoresist has a Young's modulus three orders of magnitude larger than that of the PDMS and is thus highly rigid, while Flexdym has a Young's modulus three times smaller than that of the PDMS.
In agreement with the results for the PDMS channel, the measurements for both Flexdym and SU-8 photoresist channels show 
linear dependence of $-\Delta P/Re$ on $Re$ for channels with obstacles and no dependence on $Re$ for channels without obstacles. 
These results provide additional support for the conclusion that the approximately quadratic relation between $-\Delta P$ and $Re$ for a channel with obstacles arises from inertial effects in the flow around the obstacles.

\begin{figure}[htb] 
\centering 
\includegraphics[width=0.6\textwidth]{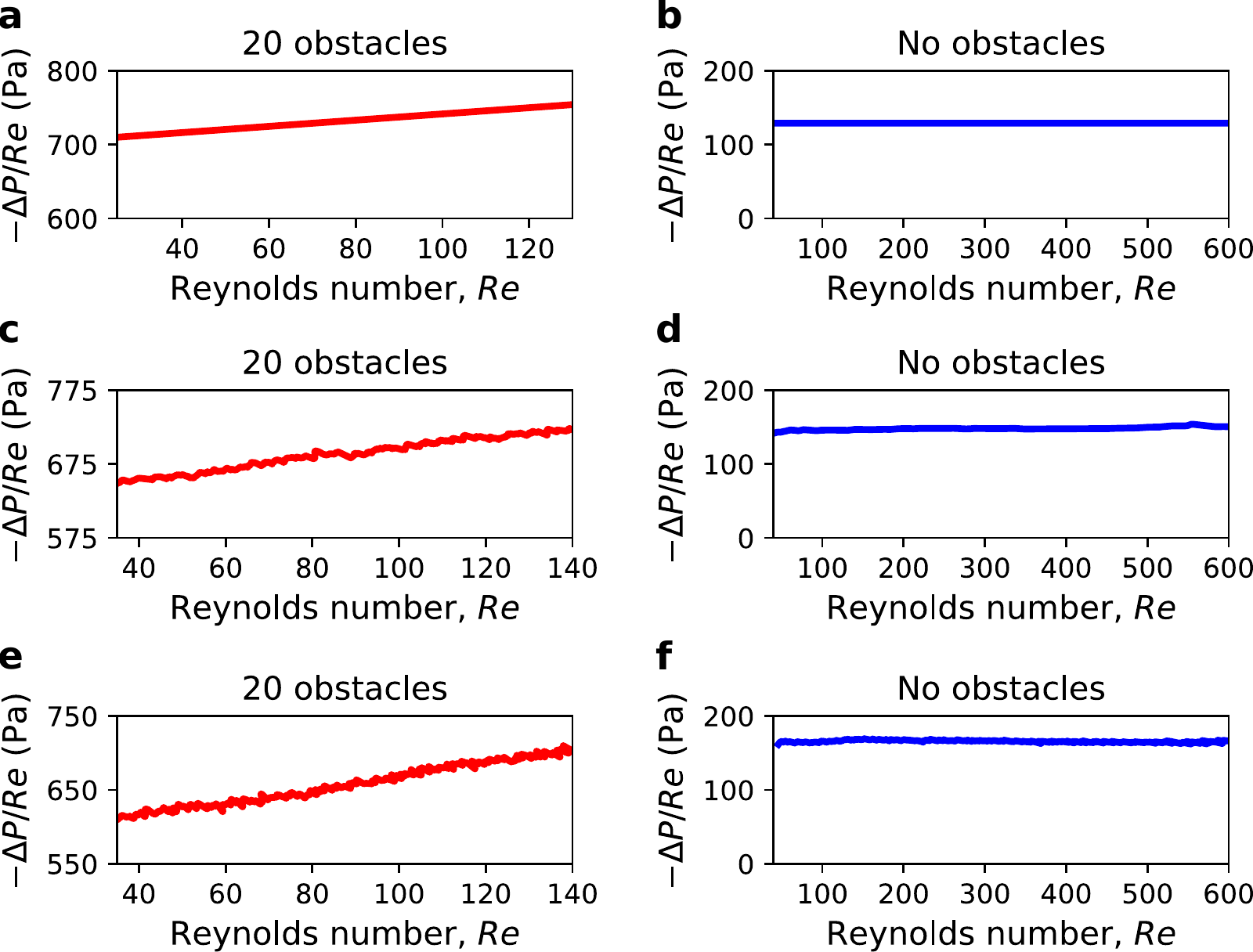}
\caption{\textbf{Nonlinear flow in Flexdym, PDMS, and SU-8 photoresist channels.}  
\textbf{a-f}
Experimental measurements of flow rate for Flexdym (\textbf{a, b}), hardened PDMS (\textbf{c, d}), and SU-8 photoresist (\textbf{e, f}) channels with (\textbf{a, c, e}) and without (\textbf{b, d, f}) obstacles. 
The range of $Re$ shown in all panels corresponds to approximately the same range of driving pressure, where the higher values of $Re$ in \textbf{b}, \textbf{d}, and \textbf{f} result from the lower hydraulic resistance in the absence of obstacles. 
The channel dimensions are the same as those used in Fig.~\ref{fig2}e-f, within the limits of experimental realization.
}\label{ExpFlex3d}
\end{figure}

We also note that the flow switching behavior  has no reliance on the manual valve used in the setup of Fig.~\ref{fig4}. We experimentally demonstrate the flow switch explicitly using a system with a linking channel without a valve in Supplementary Fig.~\ref{ExpSwitch}. 

\begin{figure}[h!!] 
\centering 
\includegraphics[width=0.5\textwidth]{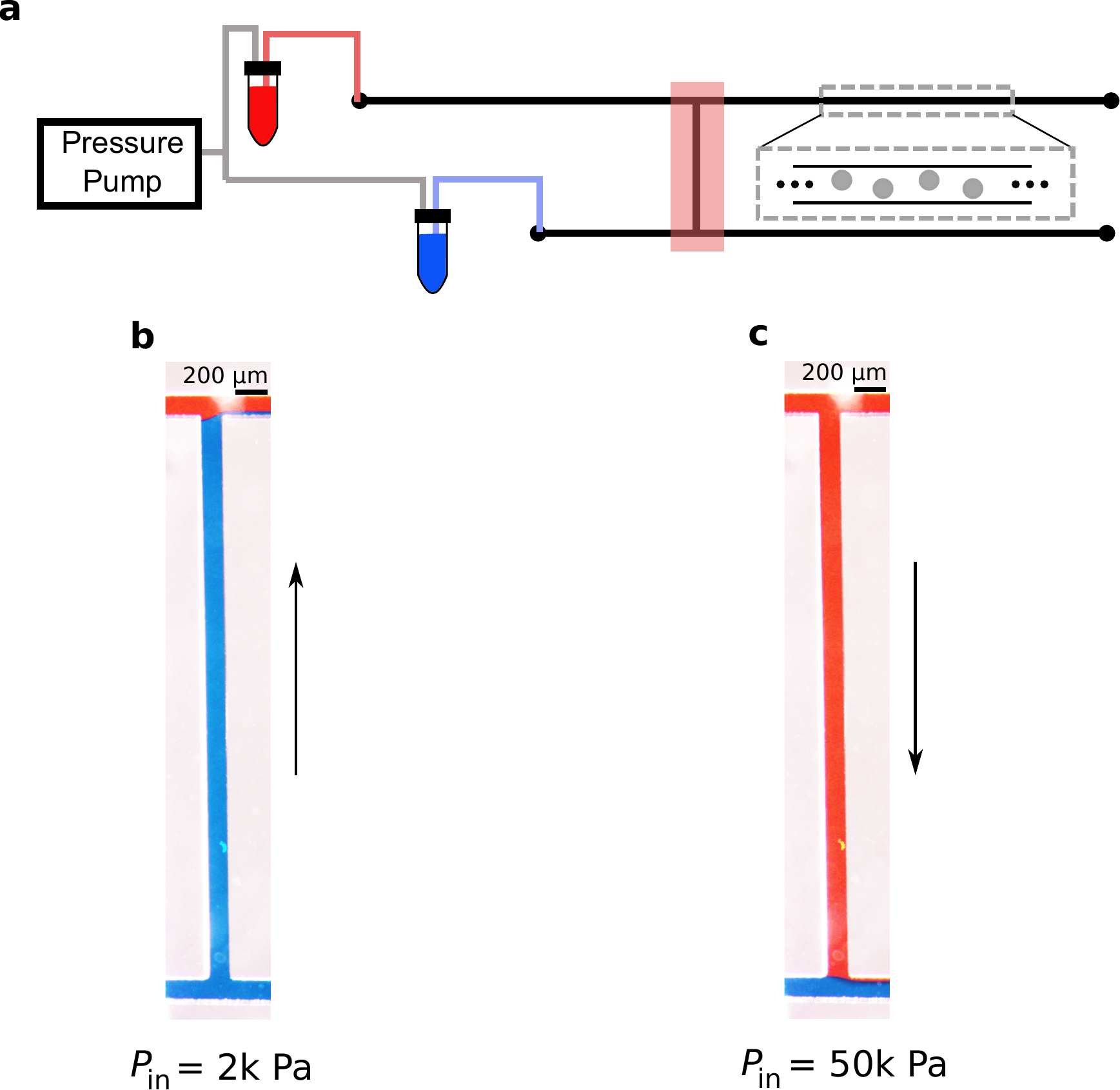}
\caption{\textbf{Experimental demonstration of flow switching.}
\textbf{a}, Schematic of microfluidic network used in Fig.~\ref{fig4} without linking channel valve.  \textbf{b}, \textbf{c}, Experimental images of flow through the linking channel for $P_{\mathrm{in}}$ below (\textbf{b}) and above (\textbf{c}) $P_{\mathrm{in}}^*$ for a camera view corresponding to the red-shaded portion in \textbf{a}, where arrows indicate flow directions. All dimensions are the same as in Fig.~\ref{fig4}, except for the linking channel, which has length $0.6\,$cm. 
}
\label{ExpSwitch}
\end{figure}

\newpage
Finally,
in Supplementary Fig.~\ref{ExpTimeSeries} we show further characterization of Braess's paradox through experimentally collected time series data for $Q_4$, $Q_5$, and $Q_4+Q_5$ as the linking channel is sequentially opened and closed for a driving pressure above $P_{\mathrm{in}}^*$. 
This supplements the results presented in Fig~\ref{fig4}b-d, where 
we present an experimental demonstration of the paradox as evidenced by the increase in the average flow rates $Q_4$ and $Q_5$ when the linking channel valve is closed. 
Supplementary Fig.~\ref{ExpTimeSeries}a,b shows clear, consistent transitions from higher to lower flow rates through both channel 4 and channel 5 each time the linking channel is opened. 

\begin{figure}[htb] 
\centering 
\includegraphics[width=0.7\textwidth]{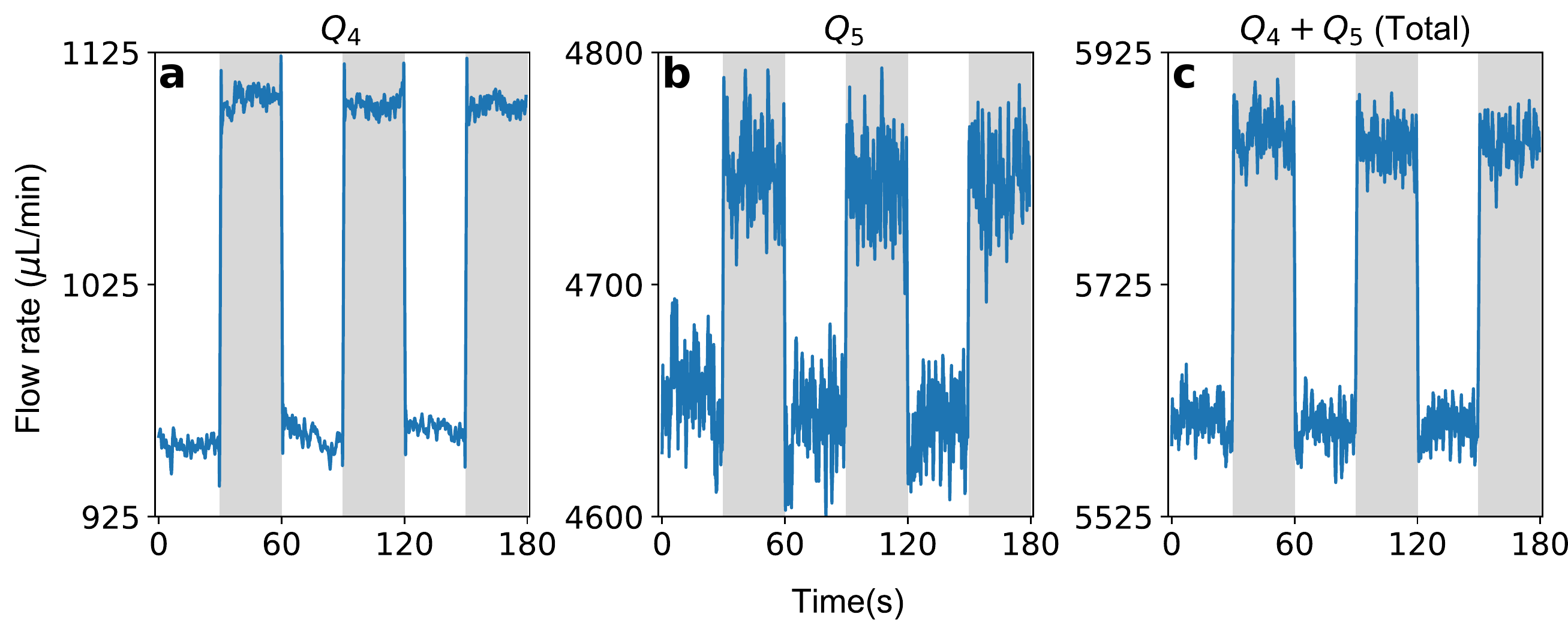}
\caption{\textbf{Experimental observation of flow rate impact of linking channel.}
\textbf{a-}\textbf{c}, Time series of the measured flow rate through channel 4 (\textbf{a}), through channel 5 (\textbf{b}), and through the combination of channels 4 and 5 (\textbf{c}). The linking channel is sequentially opened (white shade) and closed (gray shade) in 30 second intervals for $P_{\mathrm{in}}$ maintained at $80\,$kPa. All channel segment lengths are the same as in Fig.~\ref{fig4}.
All channels have a height of $220\,\mu$m and a width of $195\,\mu$m, and the obstacles have a diameter of $99\,\mu$m.
}
\label{ExpTimeSeries}
\end{figure}

\newpage
\section{Supplemental results for multiswitch networks}\label{S6}

In the following sections we outline our method for designing networks with multiple programmed switches, and we show 
examples of multiswitch networks in experiments and simulations.

\subsection{Designing multiswitch networks}\label{S6.1}
We expand on the details of how larger networks, such as those depicted in Fig.~\ref{fig1}b, can be systematically designed to exhibit multiple switches. We consider a network with multiple flow switches to be programmable if the channel dimensions can be chosen so that each individual flow switch occurs at a predefined driving pressure.

The model for a multiswitch network is constructed in the same manner as equations~(\ref{model1})-(\ref{model8}), whereby a pressure-flow relation is associated to each channel segment along with conservation equations for each of the channel junctions. For the purpose of designing multiswitch networks, we consider the pressure-flow relations for all channel segments without obstacles (including linking channels) to be of the form of equation~(\ref{eq1}). For the segments with obstacles, relations of the form of equation~(\ref{eq:5}) are used.  For the ten-switch network in Fig.~\ref{fig6}a, this amounts to $56$ equations, including three nonlinear equations corresponding to the channel segments with obstacles. For given channel dimensions, the critical switching pressure for a specific linking channel can be determined by including an additional constraint into the model that enforces the flow rate through the corresponding linking channel to be zero. The resulting set of $57$ equations can then be solved for the critical switching pressure of the linking channel. This process is repeated for each linking channel to yield the set of ten driving pressure values for which the switches occur. 

The design challenge is then to determine the channel segment dimensions such that the values of $P_{\mathrm{in}}$ at which the flow switches occur correspond to the predefined set of target pressures. If all channel dimensions are specified, the system of equations for large networks can be  solved numerically using a  root finding method or least-squares approach. Therefore, to determine the channel dimensions that achieve the set of target pressures, we define a nonlinear optimization problem whereby the adjustable parameters are a subset of the channel segment dimensions. The objective function to be minimized is a measure of the distance of the set of switching pressures from the set of target pressures. This approach can be effective in ordering the switches over the working pressure range even if the objective function cannot be brought to zero, which is of relevance  since the set of tunable channels in a network can be limited in specific applications. The approach is suitable for use in general applications, especially given that achieving the exact predefined switching pressures is expected to be less important 
in practice than having the switches occur in the specified order. 

\begin{figure}[htb] 
\centering 
\includegraphics[width=0.85\textwidth]{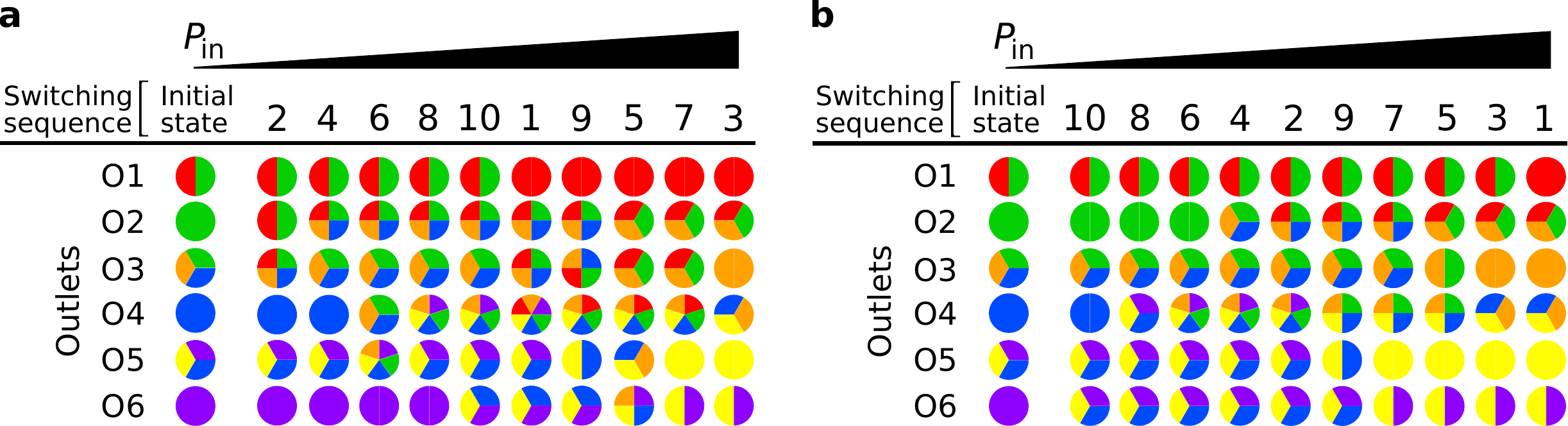}
\caption{\textbf{Alternative switching sequences for multiswitch network.} 
 \textbf{a},\textbf{b} Patterns of outlet flows achieved through optimization of the 
ten-switch network presented in Fig.~\ref{fig6} for two targeted switching sequences.
The channel segment dimensions that give rise to each sequence are presented in Supplementary Table~\ref{multiswitchTable}.
}
\label{suppSeq}
\end{figure}

 In Supplementary Table~\ref{multiswitchTable}, we present  the dimensions of all channel segments and the switching pressures of each linking channel prior to optimization for the ten-switch network in Fig.~\ref{fig6}.  In addition, we show three sets of targeted pressures (corresponding to the switching sequences in Fig.~\ref{fig6}b and Supplementary Fig.~\ref{suppSeq}) and the corresponding switching pressures and channel segment lengths found through optimization. The same initial network structure was used for all optimization runs. The specific objective function used during optimization is the sum of relative differences between the actual and target switching pressures for all linking channels in the network. Optimization was performed using a Nelder-Mead optimization algorithm and was implemented through the Python SciPy Optimize library.

In designing the multiswitch networks presented in this study, we considered network layouts in which obstacles were placed in the most downstream segment of every-other parallel channel. Each of the channel segments with obstacles were of the same length and contained the same number of obstacles, so that $\alpha$ and $\beta$ did not vary between them. In choosing the target pressures, we chose higher pressures 
for all upstream linking channels than all downstream linking channels, but considered any order otherwise. 
The resulting number of possible switching orders is still extremely large: $[(n_l/m)!]^m$ for a network consisting of $n_l$ linking channels distributed over $m$ layers. In the specific case of Fig.~\ref{fig6}, we have $n_l=10$ and $m=2$ (one upstream layer of five linking channels and one downstream layer of five linking channels), thus yielding $14,400$ distinct switching orders, each corresponding to a different internal flow state.

We note, however, that the optimization method outlined above is applicable to more general networks. They can include, for example, designs in which segments with obstacles are interspersed throughout the network or the coefficients of the Forchheimer nonlinearity are also considered to be adjustable.
While our results demonstrate that the inclusion of obstacles in only a small subset of channels can result in a very large number of flow states, an even larger number of flow switches, and thus flow states, are achievable by simply adding more linking channels to the network.

\newpage
\begin{table}[hb]
\centering
\includegraphics[width=0.8\linewidth]{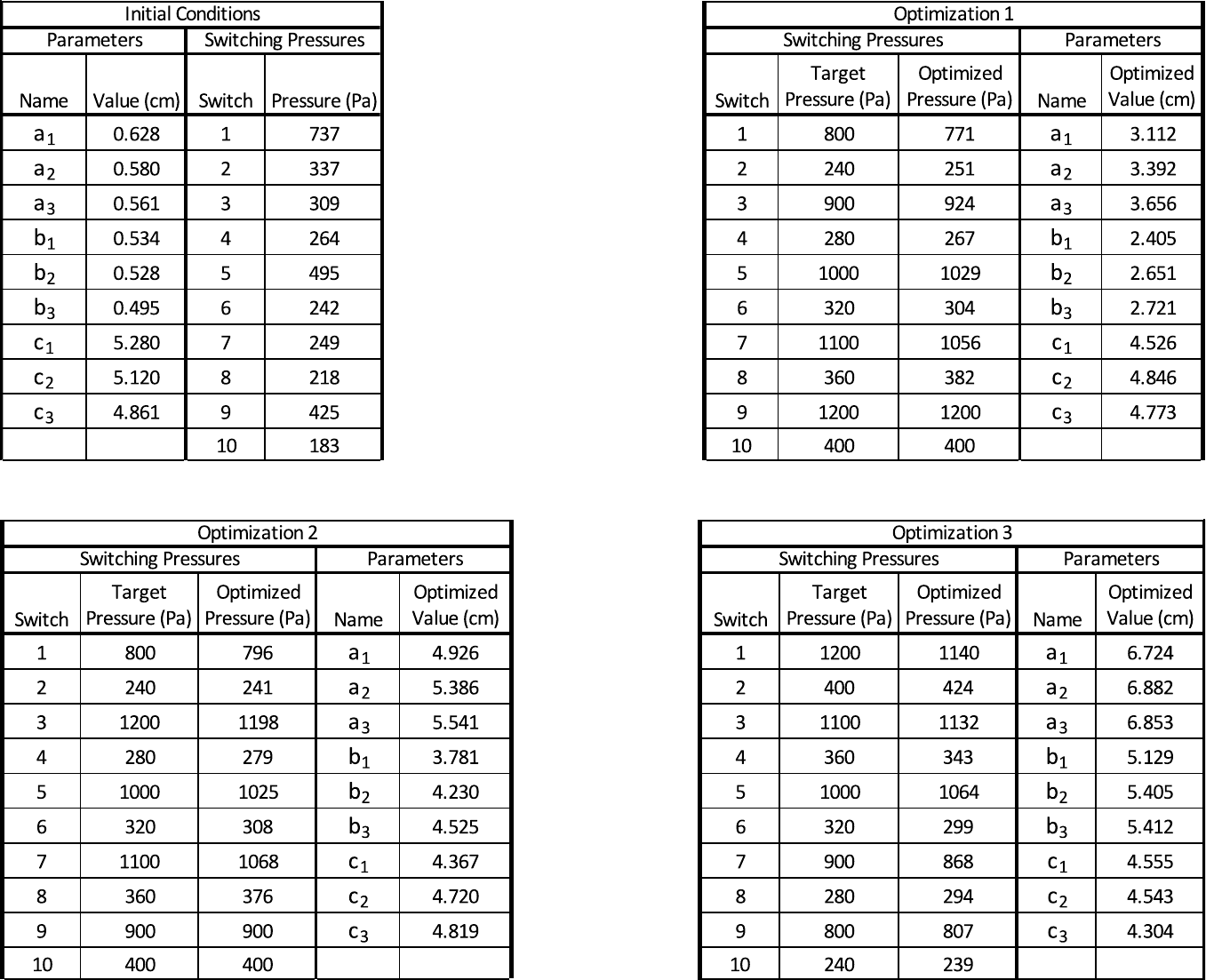}
\caption{
\textbf{Optimized design of ten-switch network.}
Top left columns: initial channel dimensions and switching pressures of the network, where the switch numbers and parameters are as marked in Fig.~\ref{fig6}a.
Top right columns: target switching pressures, optimized switching pressures, and optimized channel dimensions corresponding to the switching order in Fig.~\ref{fig6}b-d.
Bottom left columns: same as in the top right columns for the target pressures corresponding to the switching order  in Supplementary Fig.~\ref{suppSeq}a.
 Bottom right columns: same as in the top right columns for the target pressures corresponding to the switching order in Supplementary Fig.~\ref{suppSeq}b.
The other (fixed) channel segment lengths are $d=0.475\,$cm, $e=0.3\,$cm, and  $f=1.25\,$cm.
In addition, the width of the five upstream and five downstream linking channels is  $2.39\times10^{-2}$cm and $1.11\times10^{-2}$cm, respectively; the width of all other channels is $w=5\times10^{-2}$cm.
}\label{multiswitchTable}
\end{table}

\newpage
\subsection{Experimental demonstration and simulation of multiswitch network}\label{S6.2}
We have experimentally verified the switching behavior in a multiswitch network using the setup in Supplementary Fig.~\ref{ExpMultiSwitch}a, which includes six linking channels and two channel segments each containing twenty obstacles. In this experiment, dyed water is driven into each inlet by the same pressure source. Images of the flow through all channels are depicted at low (Supplementary Fig.~\ref{ExpMultiSwitch}b) and high (Supplementary Fig.~\ref{ExpMultiSwitch}c) values of the source pressure. At low driving pressure, the flows 
through the linking channels are oriented towards the channels containing obstacles. As $P_{\mathrm{in}}$ is increased, the flow direction through each linking channel switches, resulting in a pattern of flows diverging from the channels with obstacles. 

 In Supplementary Fig.~\ref{SimMultiSwitch}, we show the internal flow patterns obtained through two-dimensional simulations of a six-switch network, similar to the one used in the three-dimensional experiments. The same switching behavior is again observed as in the experiments: at low pressures, flows enter the channels with obstacles through the linking channels, and at high pressures, flows diverge from the channels with obstacles. 
This behavior is evident in both the experiments and simulations by observing the difference in the flow compositions of each outlet at low and high driving pressures. 
In particular, the flows (at the outlets)
with pure blue compositions at low pressure transition to mixed red/blue compositions at high pressure. Similarly, the flows with mixed red/blue compositions at low pressure transition to pure red compositions at high pressure.

\begin{figure}[htb] 
\centering 
\includegraphics[width=0.8\textwidth]{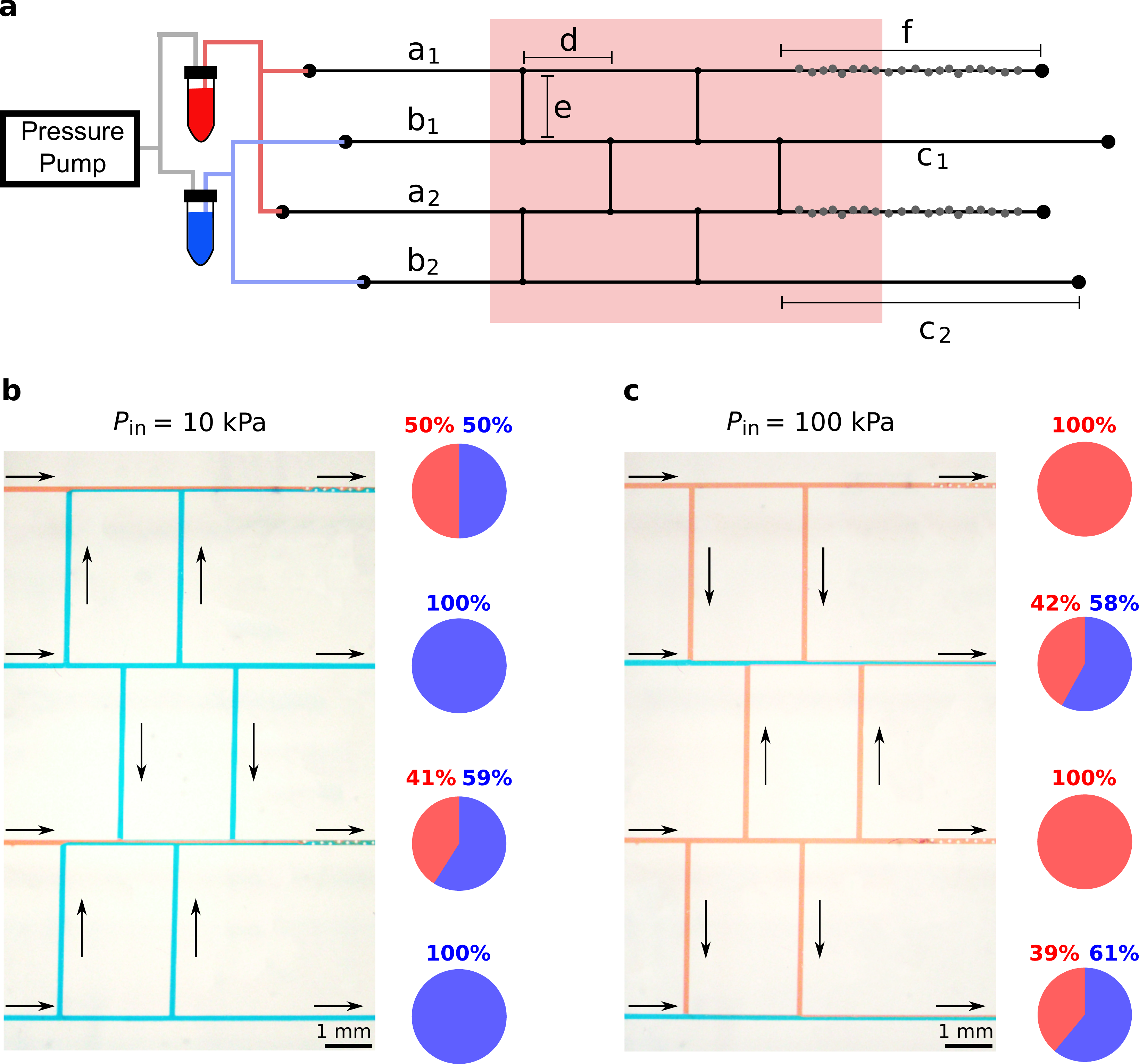}
\caption{\textbf{Experimental demonstration of six-switch network.} 
 \textbf{a}, Schematic of experimental setup, where red- and blue-dyed water is driven into individual inlets by a common pressure $P_{\mathrm{in}}$. The shaded rectangle depicts the camera view. \textbf{b}, \textbf{c}, Images of flows through the network for low (\textbf{b}) and high (\textbf{c}) driving pressure $P_{\mathrm{in}}$, where the arrows indicate flow direction. The pie charts show the flow composition at each outlet, as determined through the proportion of red and blue pixels extracted from the image along a line perpendicular to the flow direction (and upstream from the obstacles) for each outlet channel.
The dimensions of the channel segments are  $a_1=a_2=0.738, b_1=b_2 =0.515, c_1=c_2 =4.117, d=0.195, e=0.6$, and $f=1.25$ (all in units of cm). Two channel segments each contain twenty obstacles (indicated by grey circles in \textbf{a}) with diameters of $112\,\mu$m. All channels have a height of $219\,\mu$m and width of $194\,\mu$m.
}
\label{ExpMultiSwitch}
\end{figure}

\clearpage

\begin{figure}[h!!] 
\centering 
\includegraphics[width=0.8\textwidth]{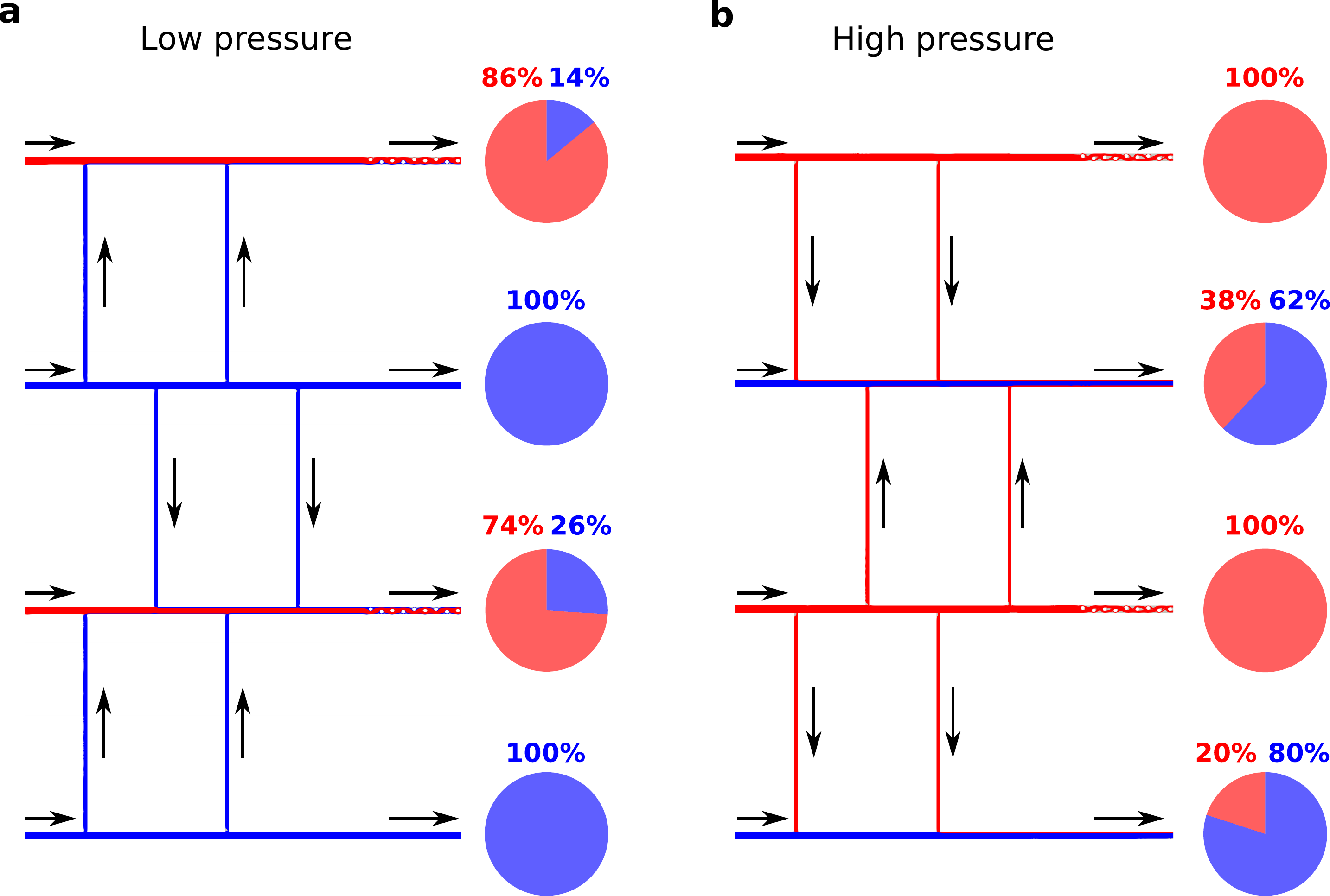}
\caption{\textbf{Simulation of six-switch network.}  
\textbf{a}, \textbf{b}, Simulation results of the Navier-Stokes equations for flow through the network presented in Supplementary Fig.~\ref{ExpMultiSwitch}a. Red and blue streamlines 
show the flows originating from the inlets with the corresponding colored fluid at low (\textbf{a}, $P_{\mathrm{in}}=1\,$kPa) and high (\textbf{b}, $P_{\mathrm{in}}=30\,$kPa) driving pressures, where the total pressure was controlled at the inlets. The view of the network corresponds to the shaded rectangle in Supplementary Fig.~\ref{ExpMultiSwitch}a and the arrows indicate the direction of flow.
The pie charts show the flow composition at each outlet, as determined through the proportion of the sample of red and blue streamlines that intersect the cross-section of each outlet channel (upstream from the obstacles).
The dimensions of the channel segments are $a_1=a_2=0.738, b_1=0.390, b_2 =0.515, c_1=c_2 =4.117, d=0.195, e=0.6$, and $f=1.05$ (all in units of cm). Two channel segments each contain twenty obstacles (indicated by grey circles in Supplementary Fig.~\ref{ExpMultiSwitch}a) with diameters of $100\,\mu$m. All linking channels have a width of $100\,\mu$m and all other channels have a width of of $200\,\mu$m.
}
\label{SimMultiSwitch}
\end{figure}

\end{document}